\documentstyle[aps,twocolumn,epsfig,amssymb]{revtex}
\begin{document}

\draft
\onecolumn
\title{Nonlinear Matter Wave Dynamics with a Chaotic Potential}

\author{S.~A.~Gardiner$^{1}$,
D.~Jaksch$^{1}$, R.~Dum$^{1,2}$,
J.~I.~Cirac$^{1}$, 
and P.~Zoller$^{1}$}
\address{$^{1}$Institut f{\"u}r Theoretische Physik,
Universit{\"a}t Innsbruck, A-6020
Innsbruck, Austria\\
$^{2}$Ecole Normale Sup\'{e}rieure, Laboratoire Kastler Brossel, 24, Rue
Lhomond, F-75231 Paris Cedex 05, France}
\date{\today}
\maketitle

\begin{abstract}
We consider the case of a cubic nonlinear Schr\"{o}dinger equation with an 
additional chaotic potential, in the sense that such a potential produces
chaotic dynamics in classical mechanics. 
We derive and describe 
an appropriate semiclassical limit to such a nonlinear Schr\"{o}dinger equation,
using a semiclassical interpretation of the Wigner function, and relate this to the
hydrodynamic limit of the Gross-Pitaevskii equation used in the context of 
Bose-Einstein condensation. We investigate
a specific example of a Gross-Pitaevskii equation with such a chaotic potential: 
the one-dimensional delta-kicked
harmonic oscillator, and its semiclassical limit. We explore the feasibility of 
experimental realization of such a
system in a Bose-Einstein condensate experiment, giving a concrete proposal of how to
implement such a configuration, and considering the problem of
condensate depletion.
\end{abstract}


\section{Introduction}
Chaos in classical Hamiltonian systems, most simply thought of as the extreme
sensitivity of trajectories in phase space to initial conditions, making long 
term predictions extremely
difficult, is by now broadly understood \cite{reichl,gutzwiller}. 
More recently, the field of quantum chaos, for our purpose meaning the study 
of quantum mechanical equivalents of classical chaotic systems, has been the 
subject of much investigation \cite{reichl,gutzwiller,haake}. From this it 
does seem that the dynamics of quantum mechanical systems can be divided into 
regular and irregular subsets, with distinct differences between the two, just 
as is the case in classical mechanics. For example, due to the unitarity of 
the evolution of the state vector, there can be no equivalent of sensitivity 
to initial conditions in the Hilbert space, but there appears to be an 
equivalent sensitivity to perturbation which distinguishes
quantum chaotic motion \cite{peres}.
A certain amount of understanding has thus been achieved, although there are still
unresolved problems, in particular how to extract classical chaos from quantum
mechanics \cite{zurek}.

Quantum dynamics are determined by the Schr\"{o}dinger equation. A
seemingly natural extension is to ask what happens when we take a quantum chaotic 
Schr\"{o}dinger equation, and add some kind of  nonlinearity. This is 
something which has been much less studied \cite{nonlinchaos}, and is certainly 
of more than academic
interest; such equations do appear in nature, for example the Gross-Pitaevskii
equation in the field of Bose-Einstein condensation \cite{bosecond,review}, 
and also in the field of
nonlinear optics \cite{nonlinearopt}. There are thus experimentally accessible 
systems in which such
chaotic effects may manifest themselves.

\begin{figure}
\setlength{\unitlength}{1mm}
\begin{center}
\begin{picture}(70,60)
\put(50,5){\framebox(25,15){\parbox{23.5mm}
{\begin{center} Hydrodynamics \end{center}}}}
\put(50,40){\framebox(25,15){\parbox{23.5mm}
{\begin{center} Classical\\ mechanics \end{center}}}}
\put(0,5){\framebox(25,15){\parbox{23.5mm}
{\begin{center} Nonlinear\\ quantum \\mechanics\end{center}}}}
\put(0,40){\framebox(25,15){\parbox{23.5mm}
{\begin{center} Quantum\\ mechanics\end{center}}}}
\put(15,27.5){\parbox{20mm}{$u\rightarrow 0$}}
\put(65,27.5){\parbox{20mm}{$u\rightarrow 0$}}
\put(32.5,15){\parbox{20mm}{$\hbar\rightarrow 0$}}
\put(32.5,50){\parbox{20mm}{$\hbar\rightarrow 0$}}
\thicklines
\put(12.5,22.5){\vector(0,1){15}}
\put(62.5,22.5){\vector(0,1){15}}
\put(27.5,12.5){\vector(1,0){20}}
\put(27.5,47.5){\vector(1,0){20}}
\end{picture}
\end{center}
\caption{Schematic diagram of how nonlinear Schr\"{o}dinger equations relate to
other forms of dynamics under various limits. The parameter $u$ represents the
strength of the nonlinearity.}
\label{connections}
\end{figure}
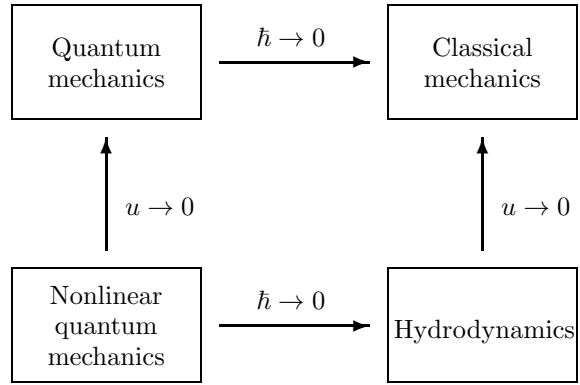

It is also interesting to note that just as in the case of quantum mechanics, where
 if
one takes the limit $\hbar\rightarrow 0$ one expects to regain classical dynamics,
one can also carry out this limit for nonlinear Schr\"{o}dinger equations. This
produces equations reminiscent of classical hydrodynamics, an 
interpretation
also extensively used in the theoretical study of Bose-Einstein condensates
\cite{hydro}. This interconnection of different kinds of dynamics is displayed
schematically in Fig.~(\ref{connections}).

We will firstly be concerned with effective ``single particle'' systems. That is to
say, where one takes a linear single particle Schr\"{o}dinger equation with a 
potential which is known to produce chaotic dynamics in Hamilton's equations of 
motion, and adds a nonlinearity to it. The Gross-Pitaevskii equation (for
example) describes the
collective dynamics of huge numbers of particles, but may nevertheless be 
thought of as an effective single particle wave equation. We later consider
corrections to this interpretation, taking into account more fully the many body
dynamics.

It is thus of general interest to determine how effects of classical chaos and
quantum chaos manifest themselves in the dynamics of nonlinear Schr\"{o}dinger 
equations, and to what extent the dynamics of the nonlinear Schr\"{o}dinger
equation can be explained by motion in the hydrodynamic limit, as determined by
the hydrodynamic equations. 

\section{Generalities}
\subsection{Gross-Pitaevskii Equation}
\label{gpeint}
In this paper we will consider explicitly only one dimensional systems, although
the analytic results presented can easily be generalized to two or three spatial
dimensions. To simplify things further, we consider only the cubic nonlinearity 
explicitly, the simplest
nonlinearity possible, resulting in the one dimensional
Gross-Pitaevskii equation, well known in the context of Bose-Einstein
condensation:
\begin{equation}
i\hbar\frac{\partial}{\partial t}\varphi=
-\frac{\hbar^{2}}{2m}\frac{\partial^{2}}{\partial x^{2}}\varphi
+V(x,t)\varphi +u|\varphi|^{2}\varphi,
\label{gpe}
\end{equation}
where $\varphi(x,t)$ is the wavefunction and $u$  the strength of the
nonlinearity. Again, the analytic results here can easily be generalized to more
complicated nonlinearities. Such a simplified system demonstrates all the main
features of a nonlinear Schr\"{o}dinger equation, and is perfectly adequate for
illustrative purposes. This kind of 
simplified system is in fact experimentally accessible,
for example in a Bose-Einstein condensate experiment, as will be shown in
Section \ref{physicalmodel}.

\subsection{Hydrodynamic Equations}
It is tempting to think of the hydrodynamic equations as the semiclassical limit
of the Gross-Pitaevskii equation. This turns out to be not quite so, as will be
shown in Sec.~\ref{wigner}. We nevertheless sketch out the standard derivation
of the hydrodynamic equations, in order to set notation, and so that later we
can point out the differences between the hydrodynamic limit and the genuine
semiclassical limit, which we will derive using Wigner functions.

We rewrite the Gross-Pitaevskii equation Eq.~(\ref{gpe}) 
using the density $\rho$ and a momentum field $P$ \cite{velocity}, 
defined in terms of the 
wavefunction
$\varphi=\sqrt{\rho}e^{iS/\hbar}$ as:
\begin{eqnarray}
\rho&=&|\varphi|^{2},\\
\rho P&=&\frac{\hbar}{2i}\left[
\varphi^{*}\frac{\partial}{\partial x}\varphi
-\left(\frac{\partial}{\partial x}\varphi^{*}\right)\varphi
\right] = \rho\frac{\partial}{\partial x}S.
\end{eqnarray}
The resulting equation of motion for the density is
\begin{equation}
\frac{\partial}{\partial t}\rho=
-\frac{\partial}{\partial x}\left(
P\rho
\right).\label{continuity}
\end{equation}
Before moving to the equation of motion for $P$, 
we first consider the equation
for $S$, which is
\begin{equation}
\frac{\partial}{\partial t}S=
-\frac{1}{2m}\left(\frac{\partial}{\partial x} S\right)^{2}
-V(x,t)- u\rho+
\frac{\hbar^{2}}{2m\sqrt{\rho}}
\frac{\partial^{2}}{\partial x^{2}}\sqrt{\rho}.
\label{prehj}
\end{equation}
The equation 
of motion for the momentum field $P$ is exactly the spatial derivative of
 Eq.~(\ref{prehj}): 
\begin{equation}
\frac{\partial}{\partial t}P=
-\frac{\partial}{\partial x}\left[
\frac{P^{2}}{2m}+
V(x,t)+u\rho-
\frac{\hbar^{2}}{2m\sqrt{\rho}}
\frac{\partial^{2}}{\partial x^{2}}\sqrt{\rho}
\right].\label{premomentumfield}
\end{equation}

Taking the hydrodynamic limit \cite{review,hydro} consists of abandoning the term in
Eq.~(\ref{premomentumfield}) proportional to $\hbar^{2}$, generally justified by
claiming that the density $\rho$ is sufficiently smooth for its derivatives
to be insignificant, resulting in
\begin{equation}
\frac{\partial}{\partial t}P=
-\frac{\partial}{\partial x}\left[
\frac{P^{2}}{2m}+
V(x,t)+u\rho
\right].\label{momentumfield}
\end{equation}
Clearly, to get Eq.~(\ref{momentumfield}), we have discarded all quantum
character of the Gross-Pitaevskii equation. Also note that if the corresponding
term is abandoned in Eq.~(\ref{prehj}) in the case where $u=0$ and $V$ is time
independent, we get
the Hamilton-Jacobi equation for a single particle in the potential $V$, with
the interpretation that $\partial S/\partial x$ is the canonically conjugate
momentum to the coordinate $x$ \cite{goldstein}. 

This seems to indicate that the hydrodynamic
equations Eqs.~(\ref{continuity},\ref{momentumfield}) might be an 
equivalent ``classical'' limit to the Gross-Pitaevskii equation with finite $u$
[Eq.~(\ref{gpe})]. As previously stated however, this turns out to be not 
quite so, as we shall soon see.

\section{Wigner Function Dynamics}
\label{wigner}
\subsection{Expansion in $\hbar$}
We wish to carry out a consistent expansion of Eq.~(\ref{gpe}) around $\hbar$,
in order to clearly separate classical from quantum dynamics, and to provide
order by order corrections. We will do this by considering the dynamics of 
the Wigner function
$W$, which is exactly equivalent to the wavefunction $\varphi$, in the sense
that all information about the wavefunction is contained within its Wigner
representation.

We define the Wigner function (for a pure state) as
\begin{equation}
W(x,p) = \frac{1}{2\pi\hbar}\int_{-\infty}^{\infty}d\tau
e^{-ip\tau/\hbar} \varphi^{*}(x-\tau/2)\varphi(x+\tau/2).
\label{wignerdef}
\end{equation}
It is well known that the dynamics of the Wigner function of a single particle 
to lowest order give
simply the classical Liouville equation of a distribution of noninteracting
particles \cite{zurek}. The exact expression to all orders in $\hbar$ for the 
time evolution
of the Wigner function $W$ is given by:
\begin{eqnarray}
\frac{\partial}{\partial t}W &=& 
\sum_{s=0}^{\infty}\frac{(-1)^{s}}{(2s+1)!} 
\left(\frac{\hbar}{2}\right)^{2s}
\frac{\partial^{2s+1}}{\partial x^{2s+1}}H
\frac{\partial^{2s+1}}{\partial p^{2s+1}}W
-\frac{\partial }{\partial p}H
\frac{\partial}{\partial x} W
\label{singleparticle}
\end{eqnarray}
where $H$ is the single particle classical Hamiltonian function. How to 
obtain this expression is sketched in Appendix \ref{wignerapp}. Setting
$\hbar=0$ we see we do indeed get the classical Liouville equation
\begin{equation}
\frac{\partial}{\partial t}W=  
\frac{\partial}{\partial x}H
\frac{\partial}{\partial p}W
-\frac{\partial }{\partial p}H
\frac{\partial}{\partial x} W,
\label{singleliouville}
\end{equation}
so long as the initial Wigner function can in fact be interpreted as a classical
probability density (i.e.\ is non-negative). If we have as a classical
Liouville density a delta
distribution, $W(x,p)=\delta(x-x_{0})\delta(p-p_{0})$,
we regain classical point dynamics. 
One can think of
a point particle being regained from quantum mechanics if we have 
a coherent state centred at $x=x_{0}$ and $p=p_{0}$ and let $\hbar\rightarrow
0$, causing the Wigner function to tend to just such a delta distribution.

It is worth mentioning that although we talk blithely about letting $\hbar$ tend
to zero, this is in fact physically meaningless. As $\hbar$ is a constant, we
must in fact expand around some scaling parameter to do with the characteristic
action scales of the problem at hand, such that at some point the quantum
corrections should be completely dominated, at least for some characteristic
time \cite{zurek}. Generally some appropriate parameter presents itself, as will
be shown in the model we present in Section \ref{model}, and expansions where it
is stated that the limit $\hbar\rightarrow 0$ is explored should be interpreted
in this manner.

What we now wish to do is to take an
equivalent limit to that presented in
Eqs.~(\ref{singleparticle},\ref{singleliouville}) 
for the Gross-Pitaevskii equation, with the object of
getting some kind of Liouville equation with the nonlinearity taken into 
account. The full expansion of the Wigner function dynamics
governed by Eq.~(\ref{gpe}) in terms of $\hbar$ turns out to be:
\begin{eqnarray}
\frac{\partial}{\partial t}W&=&
-\frac{\partial }{\partial p}H
\frac{\partial }{\partial x}W+
\sum_{s=0}^{\infty}\frac{(-1)^{s}}{(2s+1)!}
\left(\frac{\hbar}{2}\right)^{2s}
\frac{\partial^{2s+1}}{\partial x^{2s+1}}
\left[H+u\rho\right]
\frac{\partial^{2s+1}}{\partial p^{2s+1}}W,
\label{wignerexp}
\end{eqnarray}
where we have the density
\begin{equation}
\rho(x) = \int_{-\infty}^{\infty}dp'W(x,p')=|\varphi(x)|^{2},
\label{wigdens}
\end{equation}
exactly as in the hydrodynamic equations,
Eqs.~(\ref{continuity},\ref{momentumfield}). 
The result of Eq.~(\ref{wignerexp}) 
is
outlined in Appendix \ref{wignerapp}.

If we take only the zeroth term in the infinite sum, we do indeed
obtain a kind of Liouville
equation
\begin{equation}
\frac{\partial }{\partial t}W=
\frac{\partial }{\partial x}H_{\rho}
\frac{\partial }{\partial p }W
-\frac{\partial }{\partial p}H_{\rho}
\frac{\partial }{\partial x}W,
\label{liouville}
\end{equation}
where
\begin{equation}
H_{\rho}=\frac{p^{2}}{2m}+V(x,t)+u\rho,
\end{equation}
i.e.\ there is an additional ``potential'' proportional to the
density of the distribution in position space. This can be interpreted as a
large number of classical particles initially placed in phase space according to some
kind of distribution function and
interacting repulsively with one another, i.e.\ as a kind of {\em non-ideal\/}
gas. If $u$ is large we would
generally expect large numbers of such particles concentrated heavily in some
cell in position space to tend to drive one another apart, meaning that large
values of $\rho$ should in the long term be heavily disfavoured.

\subsection{Hydrodynamics Related to Wigner Function Dynamics}
\label{hydrowig}
Hydrodynamic equations can also be derived from the equation of motion for 
the Wigner function Eq.~(\ref{wignerexp}), and if one expects the hydrodynamic equations to describe a 
semiclassical limit of the Gross-Pitaevskii equation, this should be consistent with the 
semiclassical limit described by the Liouville-like equation of Eq.~(\ref{liouville}). In this section
we conclusively show this not to be the case, and explain why this is so.

In terms of the Wigner function, $P$ is
defined by:
\begin{equation}
\rho(x) P(x) = \int_{-\infty}^{\infty}dp p W(x,p),
\end{equation}
where $\rho(x)$ has already been defined by Eq.~(\ref{wigdens}). $P$ is thus
seen to be simply the first order momentum moment of the Wigner function. It
turns out to be useful to define higher order moments as well:
\begin{equation}
\rho(x) P_{n}(x) = \int_{-\infty}^{\infty}dp p^{n} W(x,p).
\end{equation}

The derivation of the equation of motion for $\rho$ is carried out in Appendix
\ref{hydroapp}, and is exactly the continuity equation of
Eq.~(\ref{continuity}), correct to all orders in $\hbar$. The equation of motion
for $P$, again to all orders in $\hbar$, turns out to be
\begin{eqnarray}
\frac{\partial}{\partial t}P&=&-
\frac{\partial }{\partial x}[V(x,t)+u\rho]
-\frac{1}{\rho m}
P_{2}(x)+
\frac{P}{\rho m}\frac{\partial}{\partial x}(\rho P) \nonumber\\
&=&-\frac{\partial}{\partial x}\left[V(x,t)+u\rho
+\frac{P_{2}}{2m}
\right]
-\frac{1}{\rho m}
\frac{\partial }{\partial x}(\sigma_{p}^{2}\rho),
\label{protohydro}
\end{eqnarray}
where
$
\sigma_{p}^{2}(x) = P_{2}(x)-P(x)^{2}
$
is the variance of the Wigner function in $p$ at a given point in $x$.

Except for the term involving $\sigma_{p}^{2}$, Eq.~(\ref{protohydro}) is
identical to the hydrodynamic equation Eq.~(\ref{momentumfield}). However, it 
can be seen that Eqs.~(\ref{continuity},\ref{protohydro}) do not form a closed
system, as the equation of motion for $P(x)$ refers to the higher order moment
$P_{2}(x)$. There is in fact, as shown in Appendix \ref{hydroapp}, an infinite
chain of differential equations for the moments $P_{n}(x)$ \cite{lill}:
\begin{eqnarray}
\frac{\partial}{\partial t}P_{n}(x)&=&\frac{P_{n}(x)}{\rho}
\frac{\partial}{\partial x}[\rho P(x)]
-\frac{1}{\rho m}\frac{\partial}{\partial x}[\rho P_{n+1}(x)]
-nP_{n-1}(x)\frac{\partial }{\partial x}[V(x,t)+u\rho]
\nonumber \\ &&
-\sum_{s=1}^{n-1}\left\{\frac{(\hbar/2)^{2s}n!}{(2s+1)![n-(s+1)]!}
P_{n-(s+1)}(x)\frac{\partial^{2s+1}}{\partial x^{2s+1}}[V(x,t)+u\rho]\right\}.
\label{motionmom}
\end{eqnarray}
In each equation the quantum corrections are described by the
sum, but there is also an
infinite chain of classical corrections; the second term of
Eq.~(\ref{motionmom}) refers to the higher order moment $P_{n+1}(x)$. To get
the  second hydrodynamic equations, Eq.~(\ref{momentumfield}), in
closed form from Eq.~(\ref{protohydro}), we must additionally 
 make the zeroth order moment 
approximation,
\begin{equation}
P_{n}(x)=P(x)^{n}.
\end{equation}
In Appendix~\ref{hydroapp} this is treated in a little more detail.

In order to reach the ``hydrodynamic limit'', it is necessary
to kill off all the quantum corrections, but there
is in fact a much more drastic approximation than only taking the limit
$\hbar\rightarrow 0$, as a whole chain of classical corrections must be abandoned at
the same time. The reason for the failure of the hydrodynamic equations as a
semiclassical limit can be seen by examining our initial reasoning more closely.
This was based partly on a
correspondence between the hydrodynamic limit of the linear Schr\"{o}dinger
equation and the equivalent Hamilton-Jacobi equation, however this also
implicitly assumes that the {\em interpretation\/} of the quantum wavefunction 
tends to a classical {\em point}. The Liouville dynamics given by
Eqs.~(\ref{singleliouville},\ref{liouville}) describe the motion of
classical {\em distributions}. As has already been mentioned, in the case of 
no nonlinearity ($u=0$) one can connect the two classical cases by considering a
distribution of the form $W=\delta(x-x_{0})\delta(p-p_{0})$, but when one is
considering a case where the dynamics are influenced by the 
density in position space 
$\rho$, this is clearly meaningless.

The correct semiclassical limit described in terms of moment equations 
is thus described by the following system:
\begin{eqnarray}
\frac{\partial}{\partial t}\rho &=&
 -\frac{1}{m}\frac{\partial}{\partial x}[\rho P(x)],\\
\frac{\partial}{\partial t}P_{n}(x)&=&\frac{P_{n}(x)}{\rho}
\frac{\partial}{\partial x}[\rho P(x)]
-\frac{1}{\rho m}\frac{\partial}{\partial x}[\rho P_{n+1}(x)]
-nP_{n-1}(x)\frac{\partial }{\partial x}[V(x,t)+u\rho],
\end{eqnarray}
where we must include every value of $n$. 
All of this is accounted for in Eq.~(\ref{liouville}).
It seems clear that
Eq.~(\ref{liouville}) is a simpler way of describing the correct classical
limit, and is almost certainly easier to integrate numerically.

The purpose of comparing a nonlinear Schr\"{o}dinger equation with its
semiclassical limit is that it explicitly
removes the wave-like or quantum behaviour, allowing us to see what there is
that is specifically ``quantum'' about the dynamics of the
 nonlinear Schr\"{o}dinger equation
under consideration.

\section{Model}
\label{model}
\subsection{The Delta-Kicked Harmonic Oscillator}
To gain insight into the general problem, it is useful
to take
a simple test system, which is (a) accessible experimentally, and (b) amenable 
to numerical attack. 
The system chosen is the one dimensional 
{\em delta-kicked harmonic oscillator},
which has been studied both classically 
\cite{classharm,stochastic,web,symmetry} and quantum
mechanically \cite{quantharm,borgonovi,frasca,hogg}. The total potential for 
the classical Hamiltonian consists of a standard harmonic potential perturbed by
a time dependent kicking potential:
\begin{equation}
V(x,t)=\frac{m\omega^{2}x^{2}}{2}+K\cos(kx)\sum_{n=-\infty}^{\infty}
\delta(t-n\tau),
\end{equation}
where $x$ is the position, $m$ is the particle
mass, $\omega$ the harmonic frequency, $K$ the kick strength, $k$ the
wavenumber, and $\tau$ the time interval between kicks. 

\subsection{Scaling}
\label{textscaling}
In our model, there are two basic
parameters: the kick strength $K$, and the strength of the
nonlinearity $u$. Additionally there is $\hbar$, which we have expanded around
in Section \ref{wigner}. The  parameters $K$ and $u$ need to be rescaled 
so that they remain equivalent in different regimes, as determined by a scaling
parameter which takes the place of $\hbar$.  In
the case of the delta-kicked harmonic oscillator there is a natural dimensionless
scaling parameter, which is $\eta$, the Lamb-Dicke parameter. 
\begin{equation}
\eta = k\sqrt{\frac{\hbar}{2m\omega}}
\label{eta}
\end{equation}
It should also
be pointed out that $\eta$ is a real physical magnitude, which really can be
adjusted in the laboratory, unlike $\hbar$. We
call the 
dimensionless kicking
strength $\kappa$, 
and the dimensionless nonlinearity strength $\upsilon$:
\begin{eqnarray}
\kappa&=&\frac{Kk^{2}}{\sqrt{2} m\omega^{2}},\label{kappa}\\
\upsilon&=& \frac{uk^{3}}{2\sqrt{2}m\omega^{2}}.\label{upsilon}
\end{eqnarray}
It is shown in Appendix \ref{scalingapp} that $\kappa$ and $\upsilon$ 
have an equivalent effect on the overall dynamics for any
value of $\eta$.

If, as is often the case when the trapping potential is harmonic, 
the Gross-Pitaevskii equation has been rescaled in
terms of harmonic coordinates $(\hat{x}_{h}=\sqrt{m\omega/\hbar}\hat{x}$,
$\hat{p}_{h}=\hat{p}/\sqrt{m\hbar\omega})$, then it can be written in terms of these
dimensionless parameters as:
\begin{eqnarray}
i\frac{\partial}{\partial t_{h}}\varphi &=&
-\frac{1}{2}\frac{\partial^{2}}{\partial x_{h}^{2}}\varphi
+V(x_{h},t_{h})\varphi +\frac{\upsilon}{\eta^{3}}|\varphi|^{2}\varphi,
\label{NLSEscaled}\\
V(x_{h},t_{h})&=&\frac{x_{h}^{2}}{2}+\frac{\kappa}{\sqrt{2}\eta^{2}}
\cos(\sqrt{2}\eta x)\sum_{n=-\infty}^{\infty}
\delta(t_{h}-n\tau_{h}).
\label{potentialscaled}
\end{eqnarray}
The wavefunctions have  been rescaled so that they are properly
normalized with respect to the harmonic position coordinate, and the 
time evolution is with respect to the dimensionless 
time $t_{h}=\omega t$. 
It is this form of the Gross-Pitaevskii equation that we use in our numerical 
simulations.

\section{Model Phase Space Dynamics}
\subsection{Classical Point Dynamics}
\label{poinsec}
The dynamics of a classical point particle in a delta-kicked harmonic potential
have been described fairly extensively elsewhere 
\cite{classharm,stochastic,web,symmetry}. 
Briefly,
we choose a value for $\tau_{h}$. For a given $\tau$ there is only one free
parameter which affects the phase space dynamics: $\kappa$.
There is a resonance condition $\tau_{h} = 2\pi r/q$ ($r/q$ is a positive
rational, where $q>2$), whereby 
there are interconnecting channels of chaotic dynamics in the
phase space \cite{web,symmetry}, the thickness of which depends on the kick
strength $\kappa$ \cite{web}. For $\kappa$ not too large, these form
an Arnol'd  stochastic web which  spreads through all of phase 
space, and has a characteristic $q$ symmetry. 
For large $\kappa$, one observes global chaos. 
Note that
Arnol'd diffusion \cite{arnold} can occur in systems of less than two
dimensions when the conditions for the KAM (Kolmogorov, Arnol'd, Moser) 
theorem \cite{reichl,kam} are not fulfilled, as is the case here 
\cite{stochastic,web,symmetry}.

Here [and also in the following numerical work on the Gross-Pitaevskii equation
Eq.~(\ref{gpe})
and Liouville equation Eq.~(\ref{liouville})]  
we consider the case where $\tau_{h}=2\pi/6$ and $\kappa=1$.
%In Fig.~\ref{poincare}(a), the unstable initial condition is where
%$\tilde{x}=\sqrt{2}\pi$ and $\tilde{p}=0$.
%, the stable initial condition where  $\tilde{x}=2\sqrt{2}\pi$ and $\tilde{p}=0$. 
The scaled position and momentum are defined as
\begin{eqnarray}
\tilde{x}&=&\frac{kx}{\sqrt{2}}=\eta x_{h},\label{dimpos}\\
\tilde{p}&=&\frac{kp}{\sqrt{2}m\omega}=\eta p_{h}.\label{dimmom}
\end{eqnarray}
These scaled variables are chosen so that the phase space dynamics of a
classical point particle described in terms of them are affected only by
$\kappa$ and 
$\tau_{h}$. As can be seen, they correspond 
exactly to the scaled harmonic position and momentum when
$\eta=1$.

\begin{figure}
\begin{center}
\epsfig{file=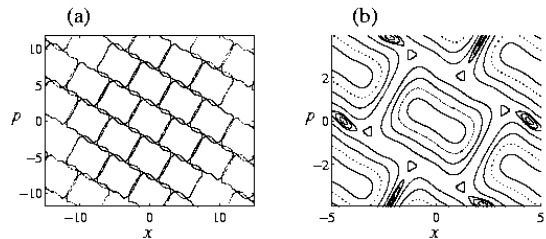,width=85mm}  
\end{center}
\caption{Poincar\'{e} sections of the phase space dynamics of the classical
delta-kicked harmonic oscillator. 
(a) Single unstable initial condition  forming a stochastic web spreading 
through phase space. 
(b) Close up of the phase space, showing the closed curves characteristic 
of regular dynamics. In both cases
$\tau_{h}=2\pi/6$, $\kappa=1$.}
\label{poincare}
\end{figure}

It can be seen in Fig.~\ref{poincare} that the phase space, in this case 
having a 6 symmetry, consists of a
stochastic web of chaotic dynamics, where an initial condition can spread 
throughout phase space,  enclosing
cells of stable dynamics. An trajectory initially inside one of these stable 
cells will generally be held in a ring of six cells, equidistant from the
centre,  for all time (with the exception
of the particle initially in the central cell, where it stays)
\cite{classharm}.

\subsection{Gross-Pitaevskii Equation}
In this section and in Sec.~\ref{liouvillesec}, we always work with the 
harmonically scaled position $x_{h}$ and momentum $p_{h}$ 
and with the dimensionless time
$t_{h}$. For the sake of brevity we omit the $h$ subscript, and thus write these
variables simply as $x$, $p$, and $t$ (or $\tau$).

We integrate numerically, using a split operator method, the
Gross-Pitaevskii equation as given in Eq.~(\ref{NLSEscaled}) considering only
the harmonic potential for periods of time of length $\tau$, 
punctuated by the exact mapping
\begin{equation}
\varphi(x,t^{+})=e^{-i\kappa\cos(\sqrt{2}\eta x)/\sqrt{2}\eta^{2}}
\varphi(x,t^{-})
\end{equation}
which accounts for the effect of the instantaneous delta kicks. 
This was carried out for various values
of $\upsilon$ and $\eta$, where $\kappa=1$ and $\tau = 2\pi/6$
in every case. 

We have calculated
the {\em time averaged\/} Wigner function, by which we mean the average of all
the Wigner functions determined just before each delta kick, for 100 kick
periods. The initial wavefunctions are displaced ground states. That is, the
ground state of the Gross-Pitaevskii equation is determined
numerically, for each value of $\upsilon$. We then locate the centre of the
wavefunction at a point which is in a regular or chaotic region of the the {\em
classical single particle\/} phase space. ``Unstable'' initial wave-packets are
centred at $x=\sqrt{2}\pi/\eta$ (harmonic units), and ``stable'' initial
wave-packets at $x=2\sqrt{2}\pi/\eta$. The initial wavefunctions are thus
centred exactly either in the middle of a cell in phase space, or in an area 
dominated by web dynamics.
These displaced states are the natural equivalent of coherent states for a cubic
nonlinear Schr\"{o}dinger equation. Just like coherent states, the density profile
keeps its shape in a simple harmonic potential as it oscillates back and forth. 
This oscillating excitation is the so-called Kohn mode \cite{kohn}. 

\begin{figure}
\begin{center}
\epsfig{file=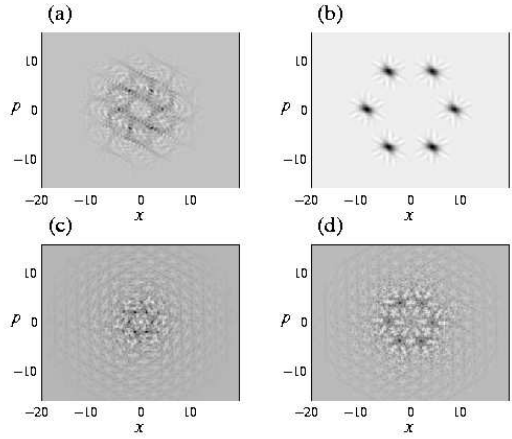,width=85mm}  
\end{center}
\caption{Pseudocolour plot of time averaged Wigner functions when 
$\upsilon=0$, i.e.\ {\em linear\/} Schr\"{o}dinger equation dynamics, 
in the two cases of: 
$\eta=1$, for (a) unstable initial condition,
(b) stable initial condition; 
$\eta=2$, for (c) unstable initial condition,
(d) stable initial condition.
Position and
momentum are scaled in harmonic units, and black means large
and positive.}
\label{uzero}
\end{figure}

Firstly we show, in Fig.~(\ref{uzero}), the case of no nonlinearity, for the
sake of reference. In this case the initial conditions are simply coherent
states. Note that because it is possible for the Wigner function to 
have negative values, the colour representing zero is in general different in
each pseudocolour plot. Thus, in each plot there is a ``background'' colour,
which represents zero, with a superimposed 
pattern made up of darker and lighter shades.
Notice that for $\eta=1$, the unstable initial condition
[Fig.~\ref{uzero}(a)] appears to move through
phase space following the stochastic web, whereas the stable initial condition
[Fig.~\ref{uzero}(b)] 
simply circles around phase space, as would an initial coherent state in a
simple harmonic potential. The wavefunction is clearly somewhat deformed (in the
case of a harmonic potential we would see perfect circles) but is otherwise well
localized and well behaved. In the case of $\eta=2$, one might be forgiven for
thinking that whether the initial condition is ostensibly stable or not is of
negligible importance. The fact that $\eta$ is larger has the effect that the
phase space is smaller compared to the size of the initial wavefunction (as
plotted here, using harmonic units), 
and also quantum corrections play a bigger role (see Appendix \ref{scalingapp}), 
leading to the
``tunneling'' seen in Fig.~\ref{uzero}(d), through classically forbidden areas
of phase space. This tunneling can take place because the eigenstates of the
Floquet operator $\hat{F}$ 
describing the period from just before one kick to just before
the next,
\begin{equation}
\hat{F}=e^{-i(\hat{x}^{2}+\hat{p}^{2})\tau/2}e^{-i\kappa\cos(\sqrt{2}\eta
\hat{x})/\sqrt{2}\eta^{2}},
\end{equation}
are highly delocalized \cite{quantharm,borgonovi,frasca}, as is described 
in Appendix~\ref{antilocalapp}.

In Fig.~\ref{utenth} equivalent plots are shown when a nonlinearity of
$\upsilon=0.1$ is added to the Gross-Pitaevskii equation.
It can be seen that this does not make very much difference to the phase space
dynamics compared to no nonlinearity (Fig.~\ref{uzero}), which is not really
unexpected.

\begin{figure}
\begin{center}
\epsfig{file=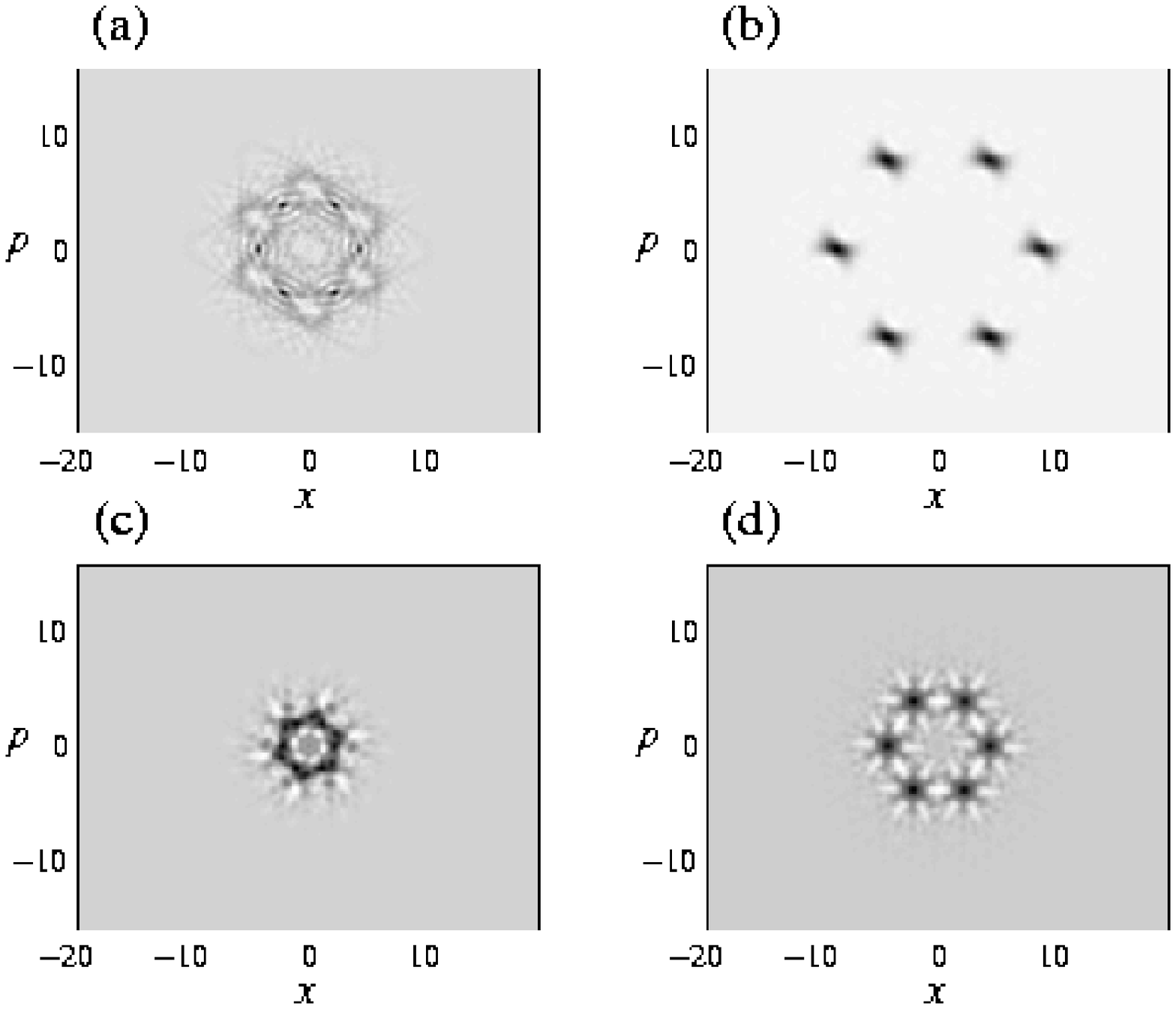,width=85mm}  
\end{center}
\caption{
As for Fig.\ref{uzero}, where $\upsilon=1$.}
\label{uone}
\end{figure}

When, as shown in Fig.~\ref{uone}, a nonlinearity of $\upsilon=1$ is added to
the Gross-Pitaevskii equation, it can be seen that this does make a difference.
Intriguingly, given that the interaction potential is more strongly repulsive, 
the phase space dynamics appear to be more strongly localized. In the case of an
unstable initial condition [Fig.~\ref{uone}(a] and \ref{uone}(c)] the web
structure is noticeably reduced, and whereas in Fig.~\ref{utenth}(d) there was
significant tunneling leading to a very delocalized phase space distribution,
in  Fig.~\ref{uone}(d) this has effectively disappeared. 

\begin{figure}
\begin{center}
\epsfig{file=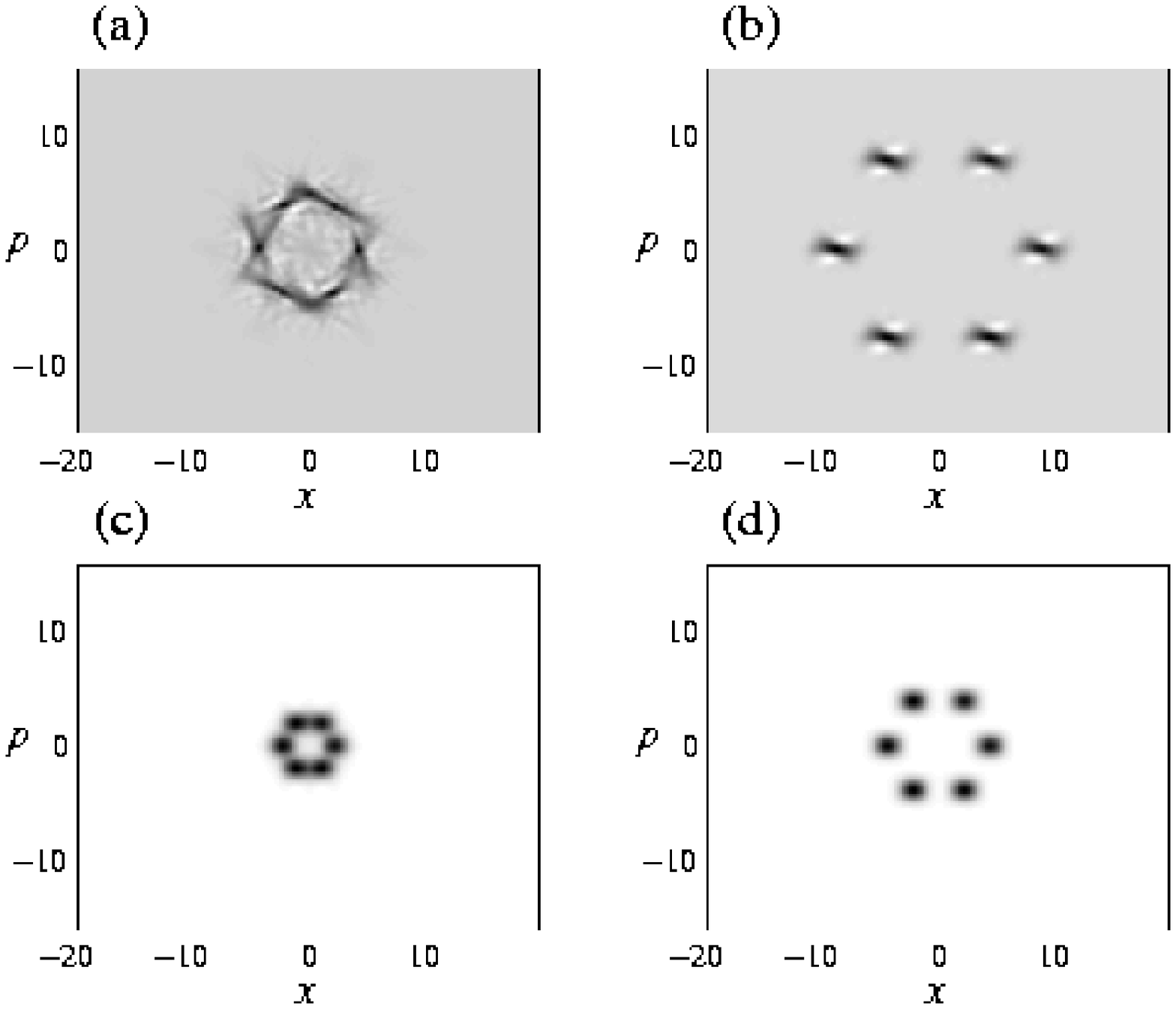,width=85mm}  
\end{center}
\caption{
As for Fig.~\ref{uzero}, where $\upsilon=10$.}
\label{uten}
\end{figure}

In Fig.~\ref{uten}, where $\upsilon=10$, this is even more marked. Where
$\eta=1$, in the case of an unstable initial condition [Fig.~\ref{uten}(a)], 
density seems to be
concentrated around a ``ring'' in phase space, based around how far out in phase
space the initial condition was. Where $\eta=2$ 
[Fig.~\ref{uten}(c,d)], whether the initial condition is ostensibly stable or
unstable, we see only six symmetrically placed round blobs of density,
analogous to a coherent state in a harmonic potential.

\subsection{Liouville Equation}
\label{liouvillesec}
Here we wish to investigate the semiclassical limit of the dynamics of the
Gross-Pitaevskii equation with a delta-kicked harmonic oscillator potential
[Eq.~(\ref{NLSEscaled})].
The appropriate dynamics are described in general by Eq.~(\ref{liouville}). As
with the Gross-Pitaevskii equation, in our case this can be carried out by
considering only the harmonic potential for periods of time $\tau$, punctuated
by an exact map describing the momentum kick.

Equation (\ref{liouville}) can be
qualitatively determined numerically by taking an ensemble of starting points
from some desired distribution, using Hamilton's equations of motion to
determine the trajectories, and using the numerically determined coarse-grained
density for the overall potential governing the motion of the individual points.
Obviously the coarse-grained density must be determined sufficiently frequently
so that between times when it is determined, it does not change enough to
have a very significant effect on the dynamics. This is in some sense analogous
to the split-step method we have used to integrate the nonlinear
Schr\"{o}dinger equation, where as the time steps shrink to length zero, the
approximate solution converges (in principle) to the exact solution. 

In each case the initial distribution is chosen by determining the ground state
of the harmonic potential Gross-Pitaevskii equation (for 
appropriate $\upsilon$ and $\eta$), shifting it so that the centre of the 
wavefunction is
at an unstable or stable fixed point (in the classical, single particle sense),
calculating the Wigner function, and interpreting this as a classical
probability distribution in $x$ and $p$. The ground state Wigner function in the
case of a harmonic potential is
always strictly nonnegative, so one can always do this.

\begin{figure}
\begin{center}
\epsfig{file=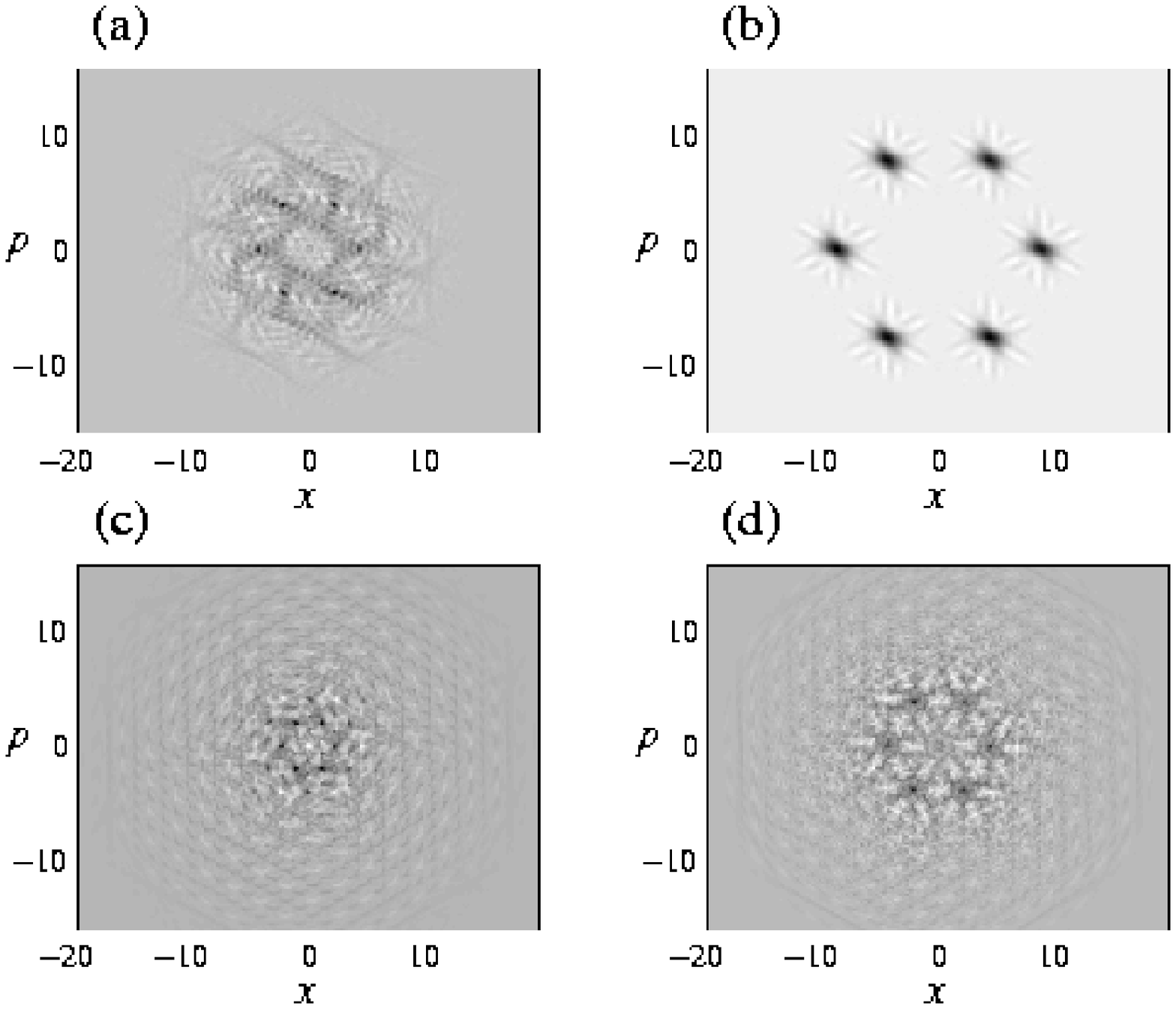,width=85mm}  
\end{center}
\caption{
As for Fig.~\ref{uzero}, where $\upsilon=0.1$.}
\label{utenth}
\end{figure}

\begin{figure}
\begin{center}
\epsfig{file=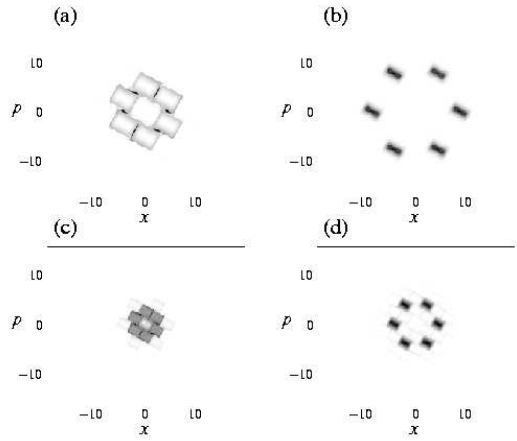,width=85mm}  
\end{center}
\caption{Pseudocolour plot of time averaged distributions undergoing Liouville
dynamics when 
$\upsilon=0.1$, in the two cases of: $\eta=1$, 
for (a) unstable initial condition, (b)
stable initial condition;
$\eta=2$, 
for (a) unstable initial condition, (b)
stable initial condition.
Black means large
and positive.}
\label{lutenth}
\end{figure}

\begin{figure}
\begin{center}
\epsfig{file=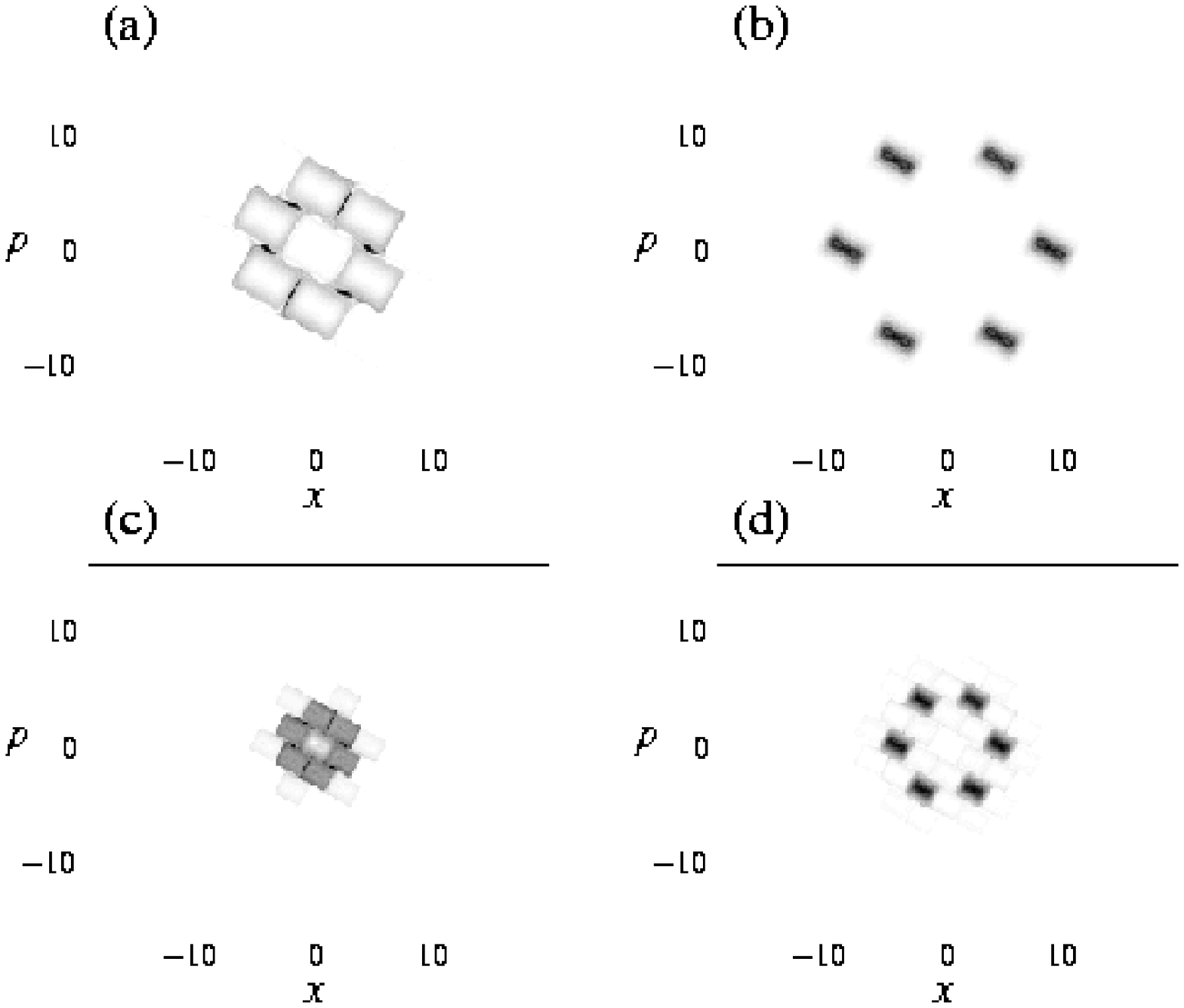,width=85mm}  
\end{center}
\caption{As for Fig.~\ref{lutenth} when 
$\upsilon=1$.}
\label{luone}
\end{figure}

\begin{figure}
\begin{center}
\epsfig{file=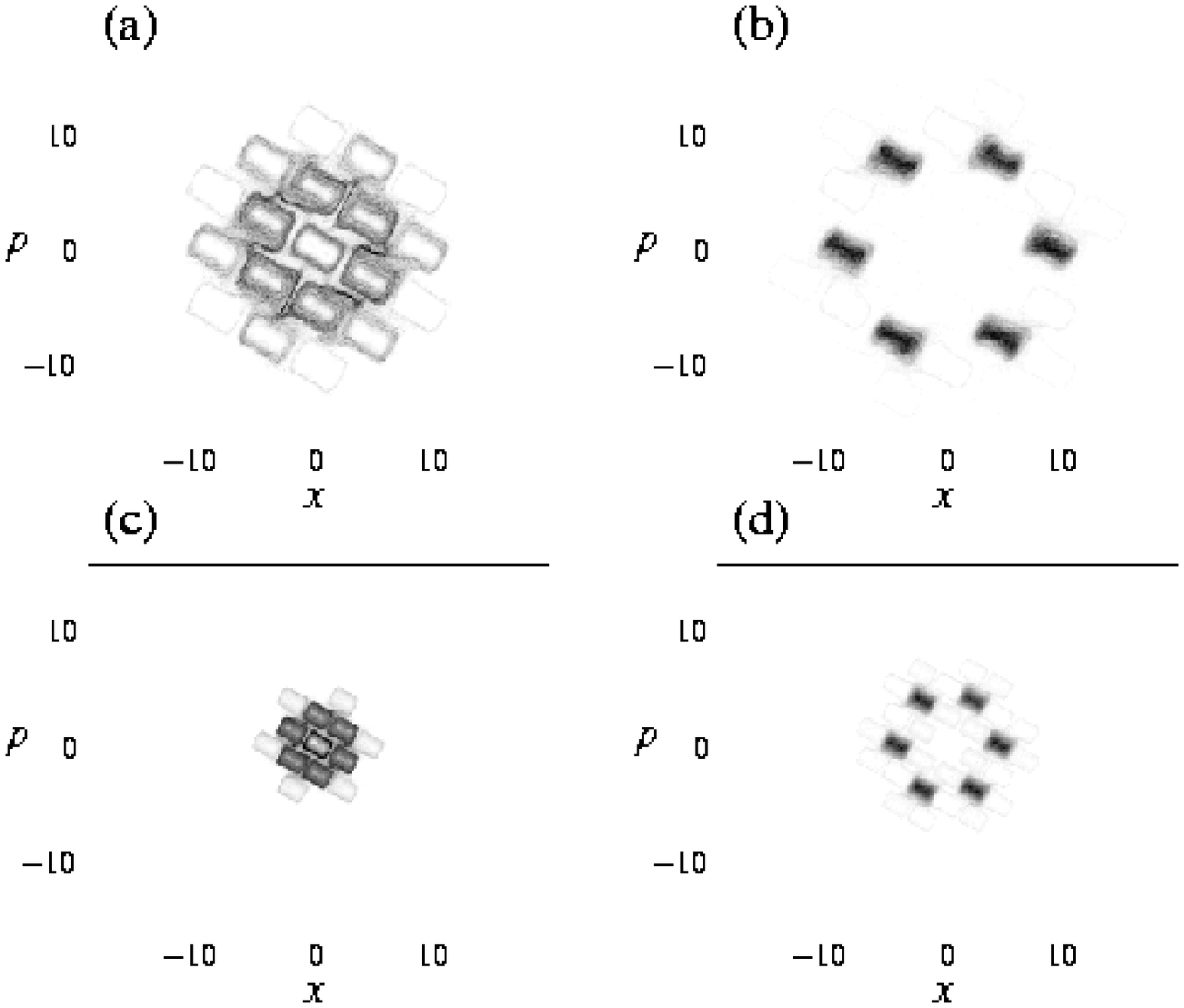,width=85mm}  
\end{center}
\caption{As for Fig.~\ref{lutenth} when 
$\upsilon=10$.}
\label{luten}
\end{figure}

Note that although $\eta$ does not enter into the dynamics of
Eq.~(\ref{liouville}) directly, by the above recipe it does enter by way of the
choice of the initial condition, which affects the effective potential due to
the distribution's density in position space, and so
on. The time averaged density distribution plots in 
Figs.~\ref{lutenth}--\ref{luten} are chosen to have
initial conditions and scaling exactly equivalent to the time-averaged Wigner
function plots shown in Figs.~\ref{utenth}--\ref{uten}.

In Fig.~\ref{lutenth} we see the density distribution averaged over 100 kicks
for the case where $\upsilon=0.1$. The dynamics are essentially similar to those
show in Fig.~\ref{poincare} for various single trajectories, and we observe a
much lesser degree of distribution through phase space when compared to the full
Gross-Pitaevskii equation (see Fig.~\ref{utenth}). In particular we see no
tunneling in Fig.~\ref{lutenth}(d), compared to Fig.~\ref{utenth}(d).
The dynamics in the cases of ``unstable'' initial conditions perhaps do not 
appear to be very strongly chaotic. Remember that only 100 kicks have been 
applied, and that in the case of the single particle classical delta-kicked harmonic
oscillator, there are {\em slow\/} chaotic  dynamics along the stochastic 
web \cite{classharm,stochastic,web}, with an overall tendency to diffuse ``outwards'' 
in phase space. We have examined the case of 100 kicks only
in order to directly compare with the the numerically determined Gross-Pitaevskii 
dynamics.

If we examine Fig.~\ref{luone}, which shows analogous dynamics to 
Fig.~\ref{lutenth} for the case that $\upsilon=1$, we observe some
increased spreading out through phase space, still contained within the
characteristic cells formed by the stochastic web in the case of the stable
initial condition for $\eta=1$. In the case of $\eta=2$ The initial distribution
seems too large for the cells, and even in the stable case there is some
diffusion outwards through phase space.

Finally we consider the case where $\upsilon = 10$, shown in Fig.~\ref{luten}.
There is significant additional diffusion through phase space for the unstable
initial condition, compared to the cases of $\upsilon=0.1$ (Fig.~\ref{lutenth})
and $\upsilon=1$ (Fig.~\ref{luone}). Even for the supposedly stable initial
condition there is some density which has found its way onto the stochastic web,
and appears to be diffusing outward. Nevertheless, the basic structure of the
single particle stochastic web appears to be retained.

There thus appears to be a clear trend, where the larger the interaction
parameter $\upsilon$, the greater the degree of diffusion outward through phase
space, but nevertheless along routes typical for single particle dynamics. 
This has a simple explanation: when $\upsilon$ is large and the distribution is
highly localized, the distribution tends to push itself apart. After this
initial explosion through phase space (actively encouraged in the unstable parts
of phase space) the contribution by the density to the effective potential is
small, and so the by now thinly spread distribution undergoes local dynamics
equivalent to single noninteracting classical particles, chaotic or stable,
depending on the location in phase space.

\subsection{Interpretation}
\subsubsection{Overview}
The most interesting thing shown by these numerical experiments, is the
conclusive demonstration that the localization observed in 
Figs.~\ref{uone}, \ref{uten} is due to interference effects, caused by terms of 
higher order in $\hbar$ in Eq.~(\ref{wignerexp}) (or more correctly,
higher order in $\eta^{2}$, as shown in Appendix.~\ref{scalingapp}). The
intuitive picture of a stronger repulsive interaction driving the Wigner
function/Liouville distribution apart, is fulfilled in the semiclassical limit,
but breaks down when all ``quantum'' corrections are accounted for.

The increasing degree of localization shown with increasing $\upsilon$ in the
Gross-Pitaevskii dynamics can also be
qualitatively explained. As is shown in Appendix \ref{antilocalapp}, in the case of
linear Schr\"{o}dinger equation dynamics, the Floquet eigenstates are highly
delocalized, due to extra symmetries connected to the fact that the wavefunction is
kicked exactly six times per oscillation period. 
The presence of delocalized eigenstates means that the wavefunction tends to spread
throughout phase space with ease; along the stochastic web if the initial condition
is in a classically unstable part of phase space, and possibly by tunneling from cell
to cell (promoted by large $\eta$) if the wavefunction is initially in a stable part
of phase space.
With increasing $\upsilon$ this
symmetry is more and more perturbed, to a point where this ability to spread freely
through phase space is lost. Interference effects due to higher order terms of the
density in
Eq.~(\ref{wignerexp}), 
%probably due to the quasi random appearance of the
%Gross-Pitaevskii wavefunction position density profile, 
act to hold the wavefunction
together, in contrast to the Liouville type dynamics described by
Eq.~(\ref{liouville}).

\subsubsection{Density in Position Space}
\begin{figure}
\begin{center}
\epsfig{file=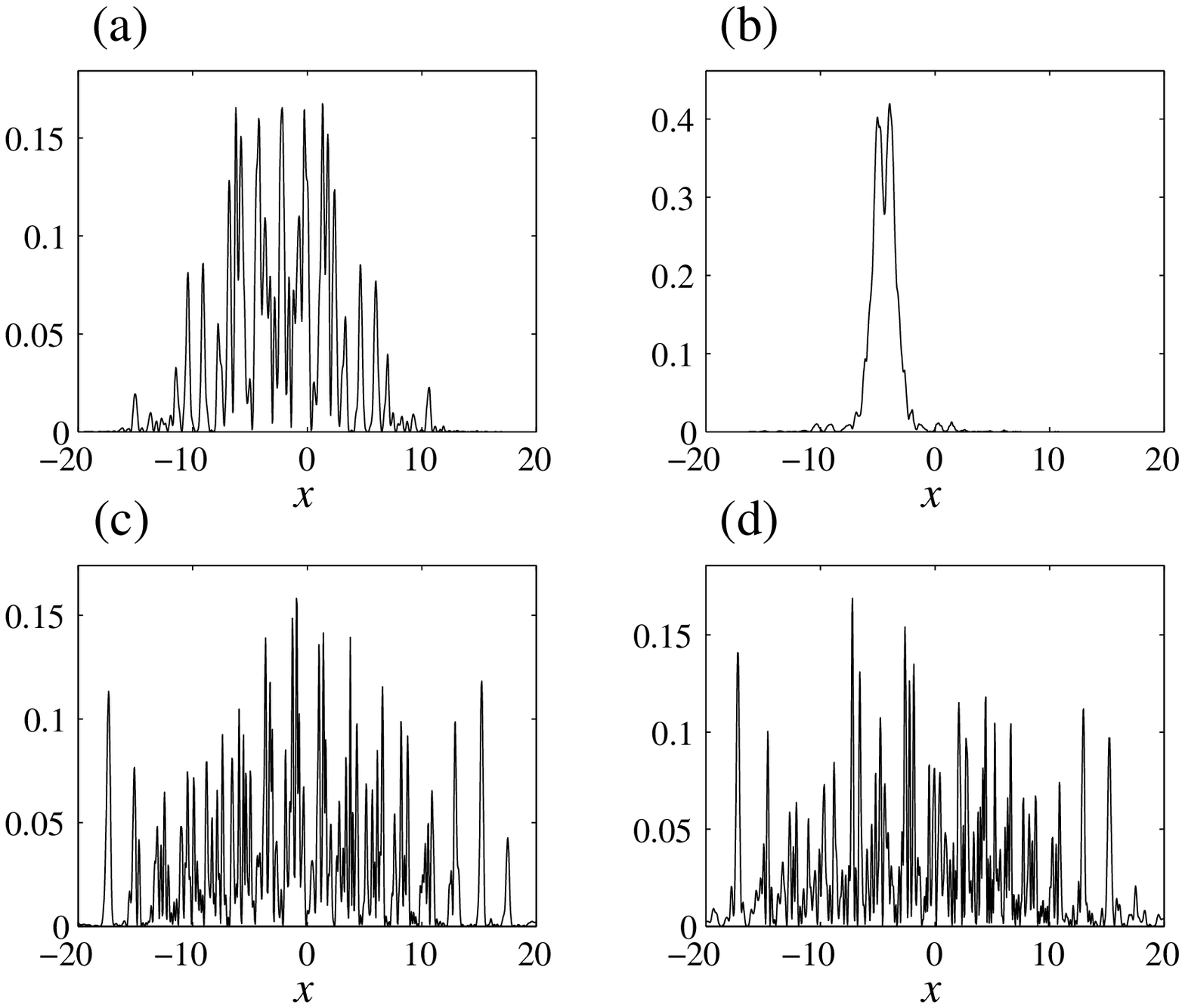,width=85mm}
\end{center}
\caption{Plots of $|\varphi(x)|^{2}$ after the application of 100 kicks and where
$\upsilon=0.1$, in the cases of:
$\eta=1$, for (a) unstable initial condition, (b) stable initial condition; and $\eta=2$,
 for (c) unstable initial condition, (d) stable initial condition.}
\label{xtenth}
\end{figure}

\begin{figure}
\begin{center}
\epsfig{file=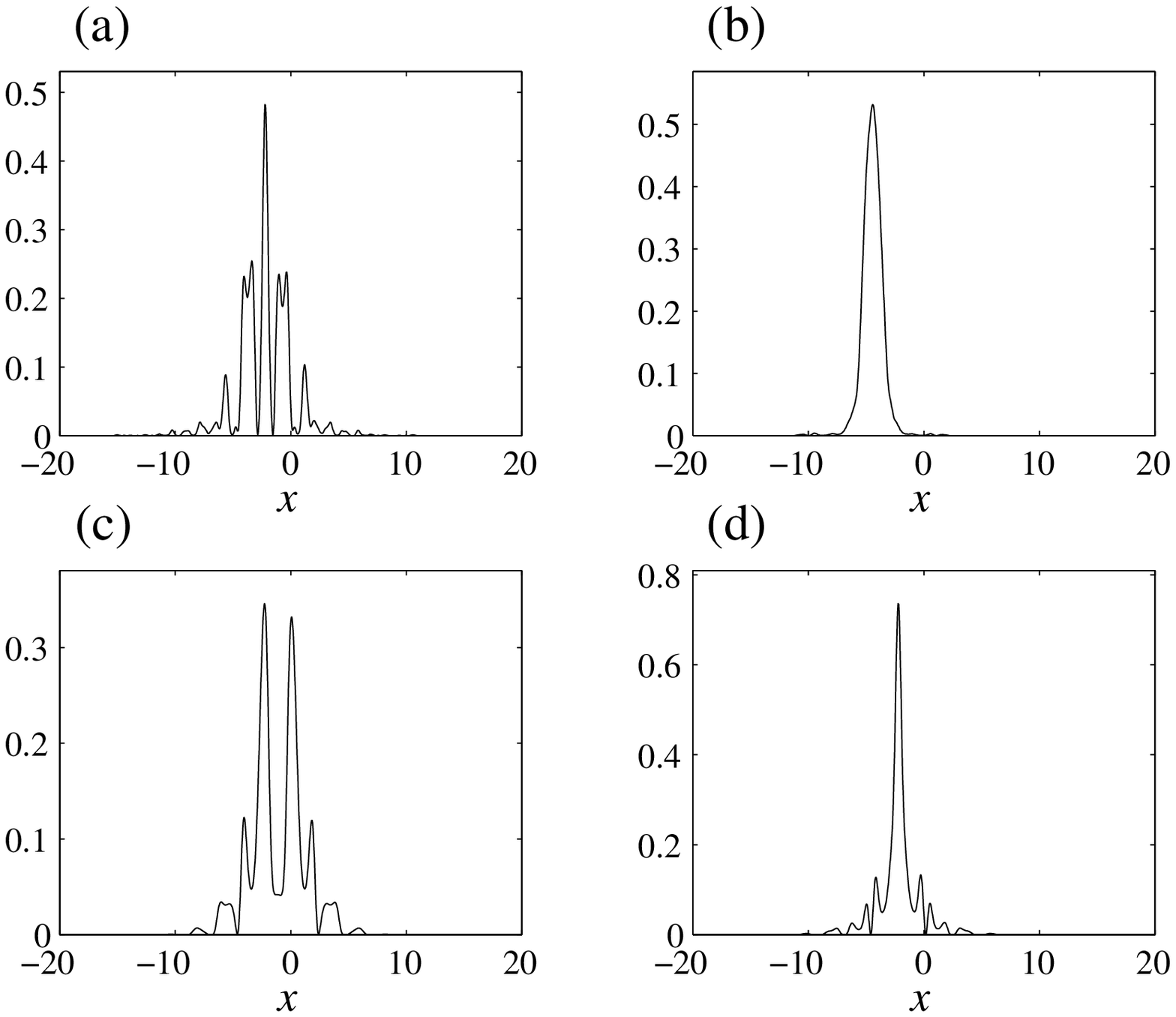,width=85mm}
\end{center}
\caption{As for Fig.~\ref{xtenth}, but for $\upsilon=1$.}
\label{xone}
\end{figure}

\begin{figure}
\begin{center}
\epsfig{file=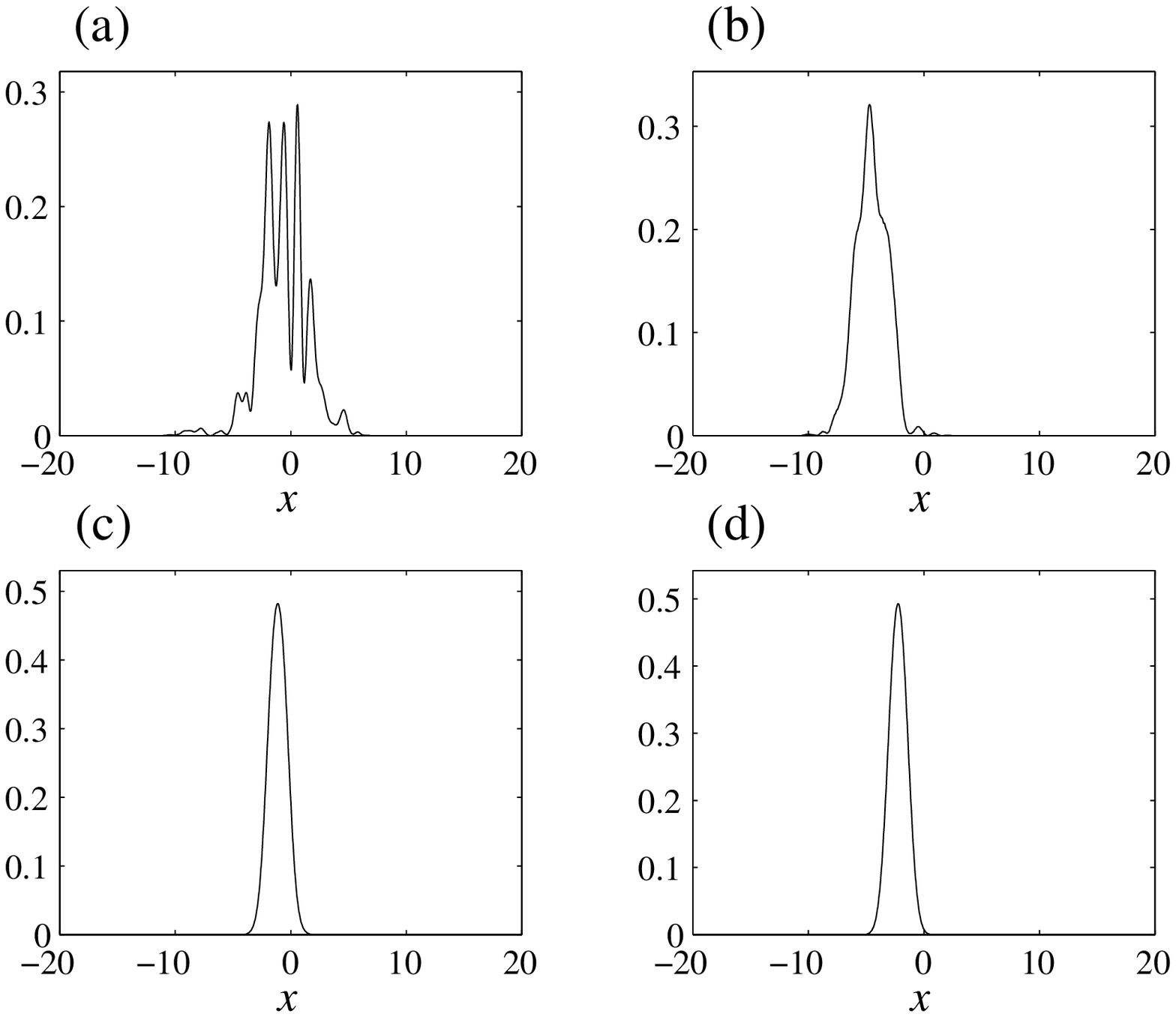,width=85mm}
\end{center}
\caption{As for Fig.~\ref{xtenth}, but for $\upsilon=10$.}
\label{xten}
\end{figure}

On this note it is instructive to look at the kinds of densities actually 
produced. We consider the final wavefunction, produced after 100 kicks, at a time just before
a hypothetical 101st kick. In Fig.~\ref{xtenth} we see plots of $|\varphi(x)|^2$ for the case 
$\upsilon=0.1$. Unsurprisingly for the unstable cases, and also for the stable case where 
$\eta=2$, the states are highly delocalized in position space, with a great deal of fine 
structure. In Fig.~\ref{xone}, this has substantially changed; the densities which were 
very complex are now much simplified, and even the stable initial condition for $\eta=1$
appears to have less structure when $\upsilon=1$ compared to $\upsilon=0.1$. When $\upsilon$
is increased to 10, as shown in Fig.~\ref{xten}, there is still some structure to the 
densities where $\eta=1$,  wheras in the case where $\eta=2$ there appears now to be none.

Obviously much more radical change is induced for the case of $\eta=2$ when increasing
$\upsilon$. Bearing in mind that $\eta^{2}$ is our effective $\hbar$, it is clear from 
Eq.~(\ref{wignerexp}) that higher order derivatives in the effective potential $V(x,t)+u\rho(x)$
will be more strongly emphasized (see also Appendix \ref{scalingapp}). Between kicks, the non-Liouville
corrections are due only to $\rho(x)$, as the derivatives of $x^{2}$ vanish.

Considering the cases of Figs.~\ref{xten}(a) and \ref{xten}(b) in particular, one might ask what 
there is about these densities which seemingly so totally dominates the dynamics. We consider the initial
state, which is simply a shifted ground state. The ground state of the Gross-Pitaevskii equation lie 
somewhere between the cases of a Gaussian (no nonlinearity) and the Thomas-Fermi limit \cite{review}, which 
is essentially an inverted parabola (large nonlinearity). With regard to the parameters we have chosen
to use, the degree of ``Thomas-Fermi-ness'' is proportional to $\upsilon/\eta^{3}$. In the 
Thomas-Fermi limit, there are no higher order derivatives of $\rho$. A Gaussian however, has an infinite
number of derivatives. For Figs.~\ref{xten}(a,b),  $\upsilon/\eta^{3}=1.25$ only.
The initial state density is thus more Gaussian than Paraboloid, and the large value of the effective 
$\hbar$ ensures that corrections due to the inevitable higher order derivatives are substantial.

Briefly:
the application of a kick scrambles the phase of the position representation of a wavefunction; instantaneously
the density in position space is unaffected however. When looking at Eq.~(\ref{wignerexp}) we see that corrections
due to higher order derivatives of $\rho$ will be emphasized for larger effective $\hbar$, in our case
$\eta^{2}$. The effect of these corrections appears to be a strong tendency for the {\em shape\/} of the
wavefunction to be preserved.

In this work, we have not really explored the regime of very large nonlinearities. In view of the fact that in the
Thomas-Fermi limit for the ground state there are no corrections to the Liouville-like equation of 
Eq.~(\ref{liouville}), it is possible that the kind of very pronounced localization observed for the case of
$\eta=2$ might again be suppressed for much larger $\upsilon$.

\subsubsection{Density in Momentum Space}
\begin{figure}
\begin{center}
\epsfig{file=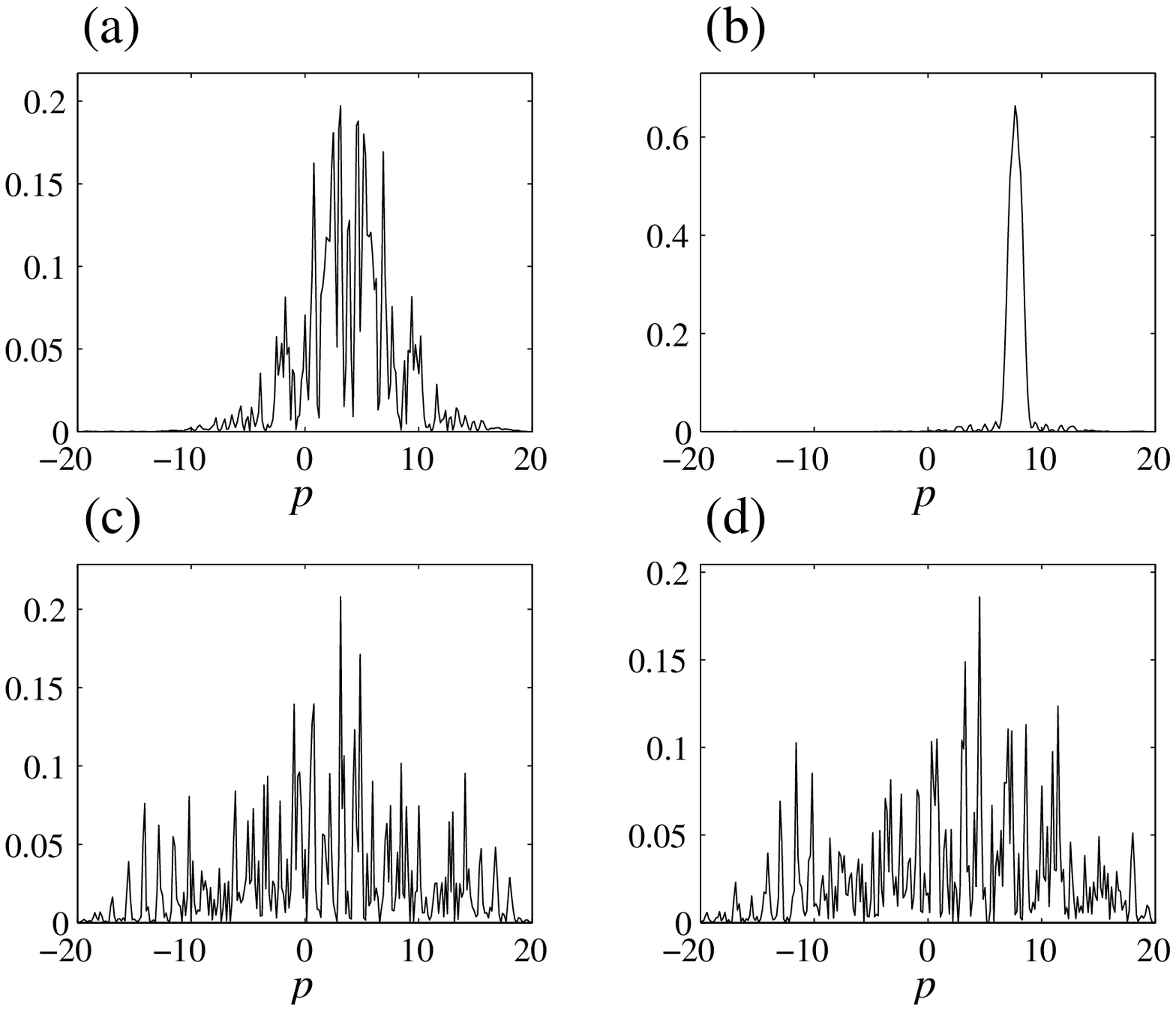,width=85mm}
\end{center}
\caption{Plots of $|\varphi(p)|^{2}$ after the application of 100 kicks and where $\upsilon=0.1$, 
in the cases of: $\eta=1$, for (a) unstable initial condition, (b) stable initial condition; and $\eta=2$,
 for (c) unstable initial condition, (d) stable initial condition.}
\label{ptenth}
\end{figure}

\begin{figure}
\begin{center}
\epsfig{file=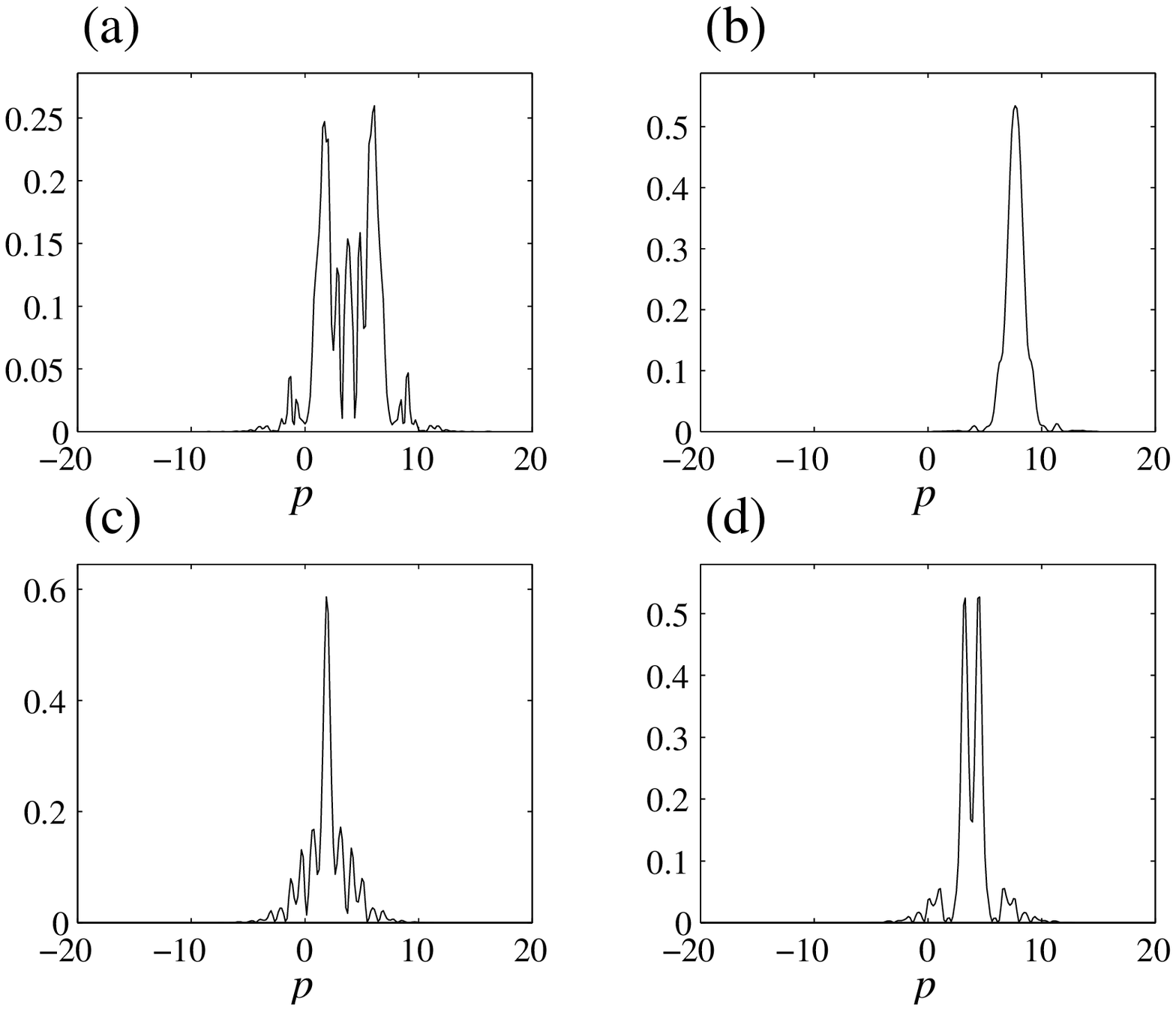,width=85mm}
\end{center}
\caption{As for Fig.~\ref{ptenth}, but for $\upsilon=1$.}
\label{pone}
\end{figure}

\begin{figure}
\begin{center}
\epsfig{file=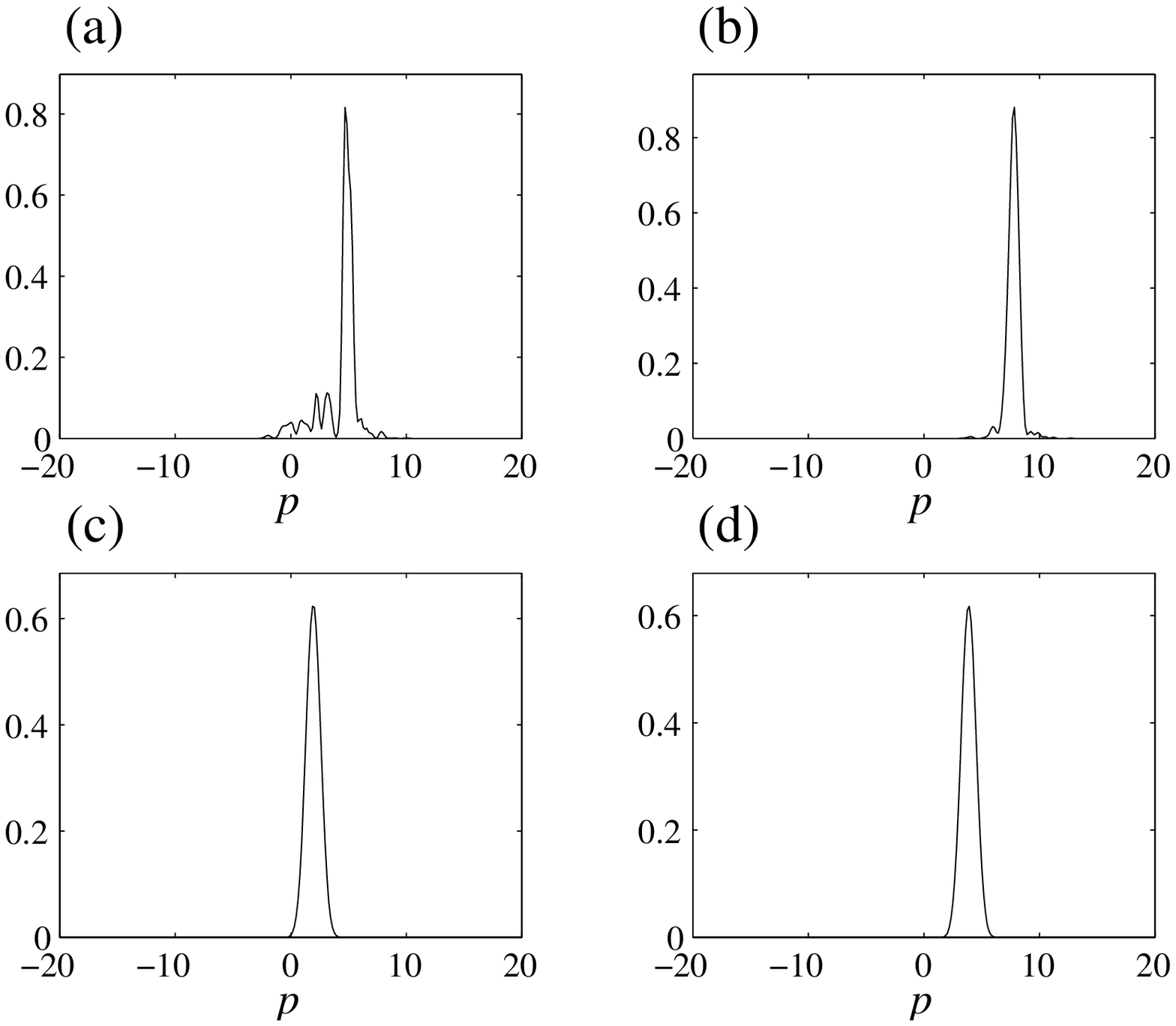,width=85mm}
\end{center}
\caption{As for Fig.~\ref{ptenth}, but for $\upsilon=10$.}
\label{pten}
\end{figure}

For the sake of comparison, in Figs.~(\ref{ptenth},\ref{pone},\ref{pten}) we show the corresponding momentum
densities to the position densities of Figs.~(\ref{xtenth},\ref{xone},\ref{xten}). The densities in position 
and momentum space essentially correspond, in that complex structure in one indicates complex structure in the other.
This is not surprising, if we consider the kinds of Wigner functions displayed in 
Figs.~(\ref{utenth},\ref{uone},\ref{uten}).

\section{Physical Model: Driven Bose-Einstein Condensate}
\label{physicalmodel}
\subsection{Introduction}
A series of pioneering experiments investigating quantum chaos with atom-optical 
systems has been carried out by Raizen and co-workers \cite{raizen}, mainly for a quantum 
realization of the delta-kicked rotor. We take a similar approach;
a possible physical realization of the delta-kicked harmonic oscillator,
consisting of a single trapped ion periodically driven by a laser, has been
described in \cite{ionchaos}. This can in principle be readily extended to a
periodically driven Bose-Einstein condensate.

\subsection{Single Particle}
We begin by regarding a single two level atom. 
In the $x$ direction, it is trapped in a harmonic potential of frequency
$\omega$, and driven time dependently by a laser field of Rabi frequency
$\Omega(t)$, wavenumber $k$, and frequency $\omega_{L}$. We disregard motional
degrees of freedom  in the
$y$ and $z$ directions as being presently uninteresting, and arrive at the 
following Hamiltonian operator:
\begin{eqnarray}
\hat{H}&=&
\frac{\hat{p}^{2}}{2m}+\frac{m\omega^{2}\hat{x}^{2}}{2}
+\frac{\hbar}{2}\{\omega_{0}(|e\rangle\langle e|-|g\rangle
\langle g|)
+\cos(k\hat{x})[\Omega(t)e^{-i\omega_{L}t}|e\rangle\langle g|+\mbox{H.c.}]\}.
\end{eqnarray}
In a rotating frame defined by 
\begin{equation}
\hat{U}=\exp[-i\omega_{L}t(|e\rangle\langle e|-|g\rangle \langle g|)/2],
\end{equation}
and in the limit of large detuning
$
|\Delta|=|\omega_{L}-\omega_{0}|\gg|\Omega(t)|
$,
$|e\rangle $ can be adiabatically eliminated to give, after transformation to an
appropriate rotating frame,
\begin{equation}
\hat{H}=
\frac{\hat{p}^{2}}{2m}+\frac{m\omega^{2}\hat{x}^{2}}{2}
+\frac{\hbar}{2}\frac{\Omega(t)^{2}}{4\Delta}
[\cos(2k\hat{x})+1]|g\rangle\langle g|.
\end{equation}
The laser is periodically switched on and off, giving a series of short 
pulses, approximated by Gaussians:
\begin{equation}
\Omega(t)^{2}=\Omega^{2}\sum_{n=-\infty}^{\infty}e^{-(t-n\tau)^{2}/\sigma^{2}},
\end{equation}
which approximate a series of delta kicks in the limit $\sigma\rightarrow 0$.
Note also that we require $\sigma\gg 1/\Delta$, otherwise  the laser is too
spectrally broad. Thus, we have finally
\begin{eqnarray}
\hat{H}&=&
\frac{\hat{p}^{2}}{2m}+\frac{m\omega^{2}\hat{x}^{2}}{2}
+\frac{\hbar\sigma\sqrt{\pi}\Omega^{2}}{8\Delta}
[\cos(2k\hat{x})+1]|g\rangle\langle g|\sum_{n=-\infty}^{\infty}\delta(t-n\tau).
\end{eqnarray}
Because we assume that the atom is always in electronic state $|g\rangle$, the
$|g\rangle \langle g|$ operator can be effectively abandoned. The extra
$+1$ 
simply adds a global phase, which can easily be accounted for,
and so this can be further simplified to:
\begin{eqnarray}
\hat{H}&=&
\frac{\hat{p}^{2}}{2m}+\frac{m\omega^{2}\hat{x}^{2}}{2}
+\frac{\hbar\sigma\sqrt{\pi}\Omega^{2}}{8\Delta}
\cos(2k\hat{x})\sum_{n=-\infty}^{\infty}\delta(t-n\tau).
\end{eqnarray}
This is exactly the Hamiltonian for the quantum delta-kicked Harmonic oscillator,
except that we have $\cos(2k\hat{x})$ instead of $\cos(k\hat{x})$. As far as
scaling is concerned, this means we must in turn consider $\eta'=2\eta$ 
instead of $\eta$ as the appropriate dimensionless parameter.

\subsection{Many Particles}
It is clear that if we consider a many particle system, then the above derivation
is independent of any particle-particle interactions which do not change the
internal states of the atoms. 
We thus consider the model 
Hamiltonian of a weakly interacting Bose gas, in second
quantized form:
\begin{equation}
\hat{H}=\int_{-\infty}^{\infty}d\vec{x}
\hat{\Psi}^{\dagger}(\vec{x})
\left[
-\frac{\hbar^{2}}{2m}\nabla^{2}+V(\vec{x},t) + 
\frac{g}{2}\hat{\Psi}^{\dagger}(\vec{x})
\hat{\Psi}(\vec{x})
\right]
\hat{\Psi}(\vec{x}),
\end{equation}
where $\hat{\Psi}
$ is the particle field operator,
$g=4\pi\hbar^{2}a_{s}/m$, and $a_{s}$ is the $s$-wave scattering length. 
We take
$V(\vec{x},t)$ to be
\begin{equation}
V(\vec{x},t)=V(x,t)+\frac{m\omega_{r}^{2}}{2}(y^{2}+z^{2}),
\end{equation}
where the potential in the $x$ direction is exactly that derived above, i.e.
\begin{equation}
V(x,t)=\frac{m\omega^{2}x^{2}}{2} +
\frac{\hbar\sigma\sqrt{\pi}\Omega^{2}}{8\Delta}
\cos(2kx)\sum_{n=-\infty}^{\infty}\delta(t-n\tau).
\end{equation}

We assume the radial frequency $\omega_{r}$ to be very large compared to the
axial frequency $\omega$ (cigar shaped
trapping configuration), and thus assume that every particle is in the harmonic
oscillator ground state in $y$ and $z$. With this assumption we can integrate
over $y$ and $z$,  reducing to a
single dimension:
\begin{equation}
\hat{H}=\int_{-\infty}^{\infty}dx
\hat{\Psi}^{\dagger}(x)
\left[
-\frac{\hbar^{2}}{2m}\frac{\partial^{2}}{\partial x^{2}}+V(x,t) + 
\frac{g_{1d}}{2}\hat{\Psi}^{\dagger}(x)
\hat{\Psi}(x)
\right]
\hat{\Psi}(x)
,
\end{equation}
where $g_{1d}=m\omega g/2\pi\hbar=2\hbar\omega_{r}a_{s}$.

\subsection{Asymptotic Expansion}
Using the particle number conserving formalism of Castin and Dum \cite{castin}, we split the 
field operator $\hat{\Psi}
$ of the many particle system into a condensate part 
and a non-condensate part:
\begin{equation}
\hat{\Psi}(x,t)=\varphi_{\mbox{\scriptsize ex}}(x,t)
\hat{a}_{\varphi_{\mbox{\tiny ex}}}(t)+\delta\hat{\Psi}(x,t),
\end{equation}
where $\varphi_{\mbox{\scriptsize ex}}$ is the {\em exact\/} condensate wave
function, and $\delta\hat{\Psi}$ describes the non-condensate particles.
Introducing the operator
\begin{equation}
\hat{\Lambda}_{\mbox{\scriptsize ex}}(x,t)=\frac{1}{\sqrt{\hat{N}}}
\hat{a}_{\varphi_{\mbox{\tiny ex}}}^{\dagger}(t)\delta\hat{\Psi}(x,t),
\end{equation}
it is possible to make asymptotic expansions of
$\hat{\Lambda}_{\mbox{\scriptsize ex}}(x,t)$, 
$\varphi_{\mbox{\scriptsize ex}}(x,t)$, such that
\begin{eqnarray}
\hat{\Lambda}_{\mbox{\scriptsize ex}}&=&\hat{\Lambda}+
\frac{1}{\sqrt{\hat{N}}}\hat{\Lambda}^{(1)}+
\frac{1}{\hat{N}}\hat{\Lambda}^{(2)}+\cdots,\\
\varphi_{\mbox{\scriptsize ex}}&=&\varphi+
\frac{1}{\sqrt{\hat{N}}}\varphi^{(1)}+
\frac{1}{\hat{N}}\varphi^{(2)}+\cdots,
\end{eqnarray}
where $\hat{N}$ is the total particle number operator.

Thus, to lowest order, the condensate particles are described by $\varphi(x)$.
The time evolution of this can be shown to be given by the Gross-Pitaevskii 
equation \cite{castin}, which in 
our case is
\begin{equation}
i\hbar\frac{\partial}{\partial t}\varphi=
-\frac{\hbar^{2}}{2m}\frac{\partial^{2}}{\partial x^{2}}\varphi+V(x,t)\varphi+
Ng_{1d}|\varphi|^{2}\varphi,
\label{physicalgp}
\end{equation}
where $N$ is the total number of particles.
In turn, the non-condensate particles are described to lowest order
by $\hat{\Lambda}(x,t)$.

The Gross-Pitaevskii equation which we have arrived at in Eq.~(\ref{physicalgp})
 can be rewritten in terms
of the dimensionless parameters $\eta'$, $\kappa$, and $\upsilon$, as described
in Sec.~\ref{textscaling}, where
\begin{eqnarray}
\eta'&=&k\sqrt{\frac{2\hbar}{m\omega}},
\label{etaprime}
\\
\kappa&=&\frac{\hbar k^{2}\sigma\sqrt{\pi/2}\Omega^{2}}{2m\omega \Delta},
\label{kappafull}
\\
\upsilon&=&\frac{8\hbar Nk^{3}\omega_{r}a_{s}}{\sqrt{2}m\omega^{2}}.
\label{upsilonfull}
\end{eqnarray}

\subsection{Non-Condensate Particles}
The mean number of the non-condensate
particles is given by $\langle \delta\hat{N} \rangle =
\langle \delta\hat{\Psi}^{\dagger} \delta\hat{\Psi}\rangle$, which to lowest
order may be described by 
$\langle \hat{\Lambda}^{\dagger}\hat{\Lambda}\rangle$. In turn, 
$\hat{\Lambda}^{\dagger}$ and $\hat{\Lambda}$ can be expanded as
\begin{equation}
\left(
\begin{array}{c}
\hat{\Lambda}(x,t)\\
\hat{\Lambda}^{\dagger}(x,t)
\end{array}
\right)
=
\sum_{k=1}^{\infty}
\hat{b}_{k}
\left(
\begin{array}{c}
u_{k}(x,t)\\
v_{k}(x,t)
\end{array}
\right)+
\sum_{k=1}^{\infty}
\hat{b}_{k}^{\dagger}
\left(
\begin{array}{c}
v_{k}^{*}(x,t)\\
u_{k}^{*}(x,t)
\end{array}
\right).
\end{equation}
which gives rise to the following equation describing the mean number of
non-condensate particles to lowest order in the perturbation expansion:
\begin{equation}
\langle \delta\hat{N}(t)\rangle = \sum_{k=1}^{\infty}
\langle \hat{b}_{k}^{\dagger}\hat{b}_{k}\rangle
\langle u_{k}(t)|u_{k}(t)\rangle +
\langle \hat{b}_{k}^{\dagger}\hat{b}_{k}+1\rangle
\langle v_{k}(t)|v_{k}(t)\rangle.
\label{noncond}
\end{equation}

The $\hat{b}_{k}$ are time-independent \cite{castin}. 
We see that the time-dependence of 
Eq.~(\ref{noncond}) is thus contained
completely within $\langle u_{k} |u_{k} \rangle$, 
$\langle v_{k} |v_{k} \rangle$. 
A system initially prepared at temperature $T$
has 
$\langle b_{k}^{\dagger} b_{k} \rangle= [\exp(E_{k}/k_{B}T)]^{-1}$, and so, if
we take the limit $T\rightarrow 0$, we get
\begin{equation}
\langle \delta\hat{N}(t)\rangle = \sum_{k=1}^{\infty}
\langle v_{k}(t)|v_{k}(t)\rangle.
\end{equation}
We thus wish to study the dynamics of $|v_{k}(t)\rangle$ to get some idea of the
change in the number of non-condensate particles, in an analogous fashion to the 
work of Castin and Dum when investigating the behaviour of a condensate held in
a time dependent isotropic harmonic potential \cite{castindumdeplete}. Note that
because the Gross-Pitaevskii equation is {\em nonlinear}, it is possible to have
chaos in the sense of exponential sensitivity to initial conditions within the
Hilbert space. If this is the case,  the above estimate of 
$\langle \delta\hat{N}(t)\rangle$ will grow automatically, due to the fact that
this estimate is essentially from a linearization around the Gross-Pitaevskii
solution \cite{castindumdeplete}. Thus the {\em rate\/} of growth of 
this estimate of $\langle \delta\hat{N}(t)\rangle$ is similar to the Lyapunov
exponent for the divergence of trajectories in phase space for discrete 
classical systems.

The dynamics of the
$|u_{k}(t)\rangle$ and 
$|v_{k}(t)\rangle$ are given by
\begin{equation}
i\hbar\frac{d}{dt}
\left(
\begin{array}{c}
|u_{k}(t)\rangle\\
|v_{k}(t)\rangle
\end{array}
\right)=
{\cal L}(t)
\left(
\begin{array}{c}
|u_{k}(t)\rangle\\
|v_{k}(t)\rangle
\end{array}
\right),
\label{Luveq}
\end{equation}
where
\begin{equation}
{\cal L}(t)=
\left(
\begin{array}{cc}
\hat{H}_{\mbox{\scriptsize GP}}(t)
+ Ng_{\mbox{\scriptsize 1d}}\hat{Q}(t)|\varphi(\hat{x},t)|^{2}\hat{Q}(t)
& Ng_{\mbox{\scriptsize 1d}}\hat{Q}(t)\varphi(\hat{x},t)^{2} \hat{Q}^{*}(t)\\
-Ng_{\mbox{\scriptsize 1d}}\hat{Q}^{*}(t)\varphi(\hat{x},t)^{*2}\hat{Q}(t)
& -\hat{H}_{\mbox{\scriptsize GP}}
-Ng_{\mbox{\scriptsize 1d}}\hat{Q}^{*}(t)|\varphi(\hat{x},t)|^{2}\hat{Q}^{*}(t)
\end{array}
\right)
\label{CasDumL}
\end{equation}
and where we have defined the Gross-Pitaevskii ``Hamiltonian'',
\begin{equation}
\hat{H}_{\mbox{\scriptsize GP}}(t)=\frac{\hat{p}^{2}}{2m}+V(\hat{x},t)+ 
u |\varphi(\hat{x},t)|^{2}-\xi(t).
\end{equation}
The phase factor $\xi(t)$ is equal to the ground state chemical potential $\mu$
when $\varphi(x,t)$ is the Gross-Pitaevskii equation ground state, for a
harmonic potential.
The projection operators $\hat{Q}$, $\hat{Q}^{*}$ are given by
\begin{eqnarray}
\hat{Q}&=&%{\mathbbm{1}}
1-|\varphi\rangle \langle \varphi|,
\\
\hat{Q}^{*}&=&%{\mathbbm{1}}
1-|\varphi^{*}\rangle \langle \varphi^{*}|,
\end{eqnarray}
where $|\varphi^{*}\rangle$ is defined by 
$\langle x|\varphi^{*}\rangle=\varphi^{*}(x)=\langle \varphi|x\rangle$.

\subsection{Dynamics of $\langle \delta\hat{N}(t)\rangle$}
To determine how $\langle \delta\hat{N}(t)\rangle$ changes over time we need to
determine the dynamics of $|v_{k}(t)\rangle$, which are coupled to the dynamics
of $|u_{k}(t)\rangle$ through Eq.~(\ref{Luveq}). We thus need to integrate
Eq.~(\ref{Luveq}), and
to integrate Eq.~(\ref{Luveq}), we need as initial conditions
$|u_{k}(0)\rangle$, $|v_{k}(0)\rangle$.

The initial conditions $|u_{k}(0)\rangle$, $|v_{k}(0)\rangle$, for 
$\varphi(x)$ in the 
ground state for a harmonic potential
are determined by diagonalizing ${\cal L}$ where $\varphi(\hat{x},t)$ is chosen
to correspond to the Gross-Pitaevskii equation ground state, for a harmonic
potential, and $\xi(t)=\mu$. For this we need to determine the
ground state condensate wavefunction $\varphi(x)$ and the ground state chemical
potential $\mu$. This is achieved by propagating the Gross-Pitaevskii equation
in imaginary time, where we use a split-operator method. 

We then determine ${\cal L}$ in the position representation
where $\varphi(x,t)$ is the previously determined
ground state and $\xi(t)=\mu$. We use a Fourier grid \cite{fouriergrid} 
to describe $\hat{p}^{2}$ in the position representation. We then diagonalize
${\cal L}$ numerically, and gain as the resultant set of eigenvectors
\begin{equation}
\left\{
\left(
\begin{array}{c}
u_{k}(x)\\
v_{k}(x)
\end{array}
\right),
\left(
\begin{array}{c}
v_{k}^{*}(x)\\
u_{k}^{*}(x)
\end{array}
\right),
\left(
\begin{array}{c}
\varphi(x)\\
0
\end{array}
\right),
\left(
\begin{array}{c}
0\\
\varphi^{*}(x)
\end{array}
\right)
\right\},
\end{equation}
with eigenvalues
$\{ E_{k}, -E_{k}, 0 , 0\}$, respectively \cite{castin}. 
These eigenvectors must be properly
normalized \cite{castin}, so that
\begin{equation}
\int_{-\infty}^{\infty}dx u_{k}^{*}(x)u_{k'}(x) -
\int_{-\infty}^{\infty}dx v_{k}^{*}(x)v_{k'}(x)=\delta_{kk'}
\end{equation}

Our initial condition for the Gross-Pitaevskii equation is in general a shifted
ground state, that is, we take the ground state wavefunction, and
instantaneously translate it in position space, otherwise altering nothing.
Physically, this could be achieved by almost instantaneously translating the
centre of the harmonic potential, so that $x^{2}\rightarrow (x-a)^{2}$.
Instantaneously, this would leave the Gross-Pitaevskii wavefunction and the
$u_{k}(x)$, $v_{k}(x)$ modes unchanged.
If we then re-express everything in terms of $x'=x-a$, we end up with the same
{\em equations\/} in terms of $x'$ as we had initially in terms of $x$, but the
{\em wavefunctions} are transformed: 
$
\{ \varphi(x), u_{k}(x),v_{k}(x)\}\rightarrow
 \{ \varphi(x'+a), u_{k}(x'+a),v_{k}(x'+a)\}
$.

Thus, if the initial Gross-Pitaevskii wavefunction is simply a shifted ground
state, then the appropriate initial $u_{k}$, $v_{k}$ are correspondingly shifted
from those determined from ${\cal L}$ for the ground state condensate
wavefunction. This set of initial conditions is in fact somewhat special;
as previously mentioned, the density profile of $\varphi(x)$ remains unchanged as it
oscillates back and forth (without kicks), the same is also true of
$u_{k}(x)$ and  $v_{k}(x)$.

Once we
have the initial conditions we can start integrating Eq.~(\ref{Luveq}).

\subsection{Numerical Results}
\label{vkevol}
\begin{figure}
\begin{center}
\epsfig{file=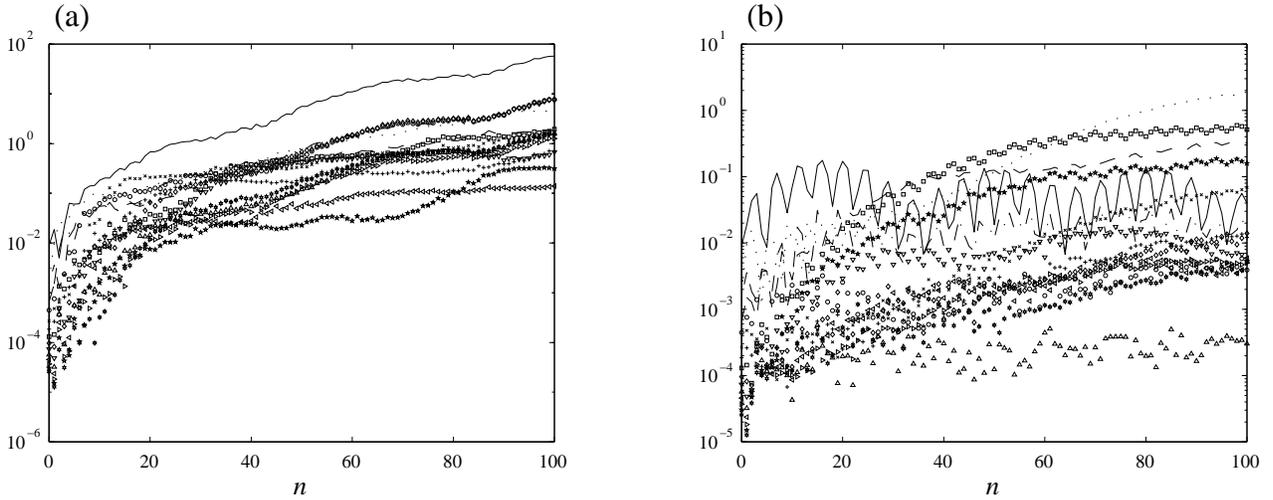,width=170mm}  
\end{center}
\caption{Semilog plot of change in $\langle v_{k}|v_{k}\rangle$ with respect 
to 
the number of kicks $n$,
for $k=1,\ldots,15$:
$k=1$ solid line,
$k=2$ dotted line,
$k=3$ dashed-dotted line,
$k=4$ dashed line,
$k=5$ circles,
$k=6$ crosses,
$k=7$ pluses,
$k=8$ squares,
$k=9$ diamonds,
$k=10$ downward pointing triangles,
$k=11$ upward pointing triangles,
$k=12$ left pointing triangles,
$k=13$ right pointing triangles,
$k=14$ pentagrams,
$k=15$ hexagrams, where
$\eta'=1$, and $\upsilon=1$. 
(a) shows data for the  ``unstable'' initial condition, where the leading term
after 100 kicks is for $k=1$, 
(b) shows data for the ``stable'' initial 
condition, where the leading term corresponds to $k=2$.}
\label{vkkoneone}
\end{figure}

\begin{figure}
\begin{center}
\epsfig{file=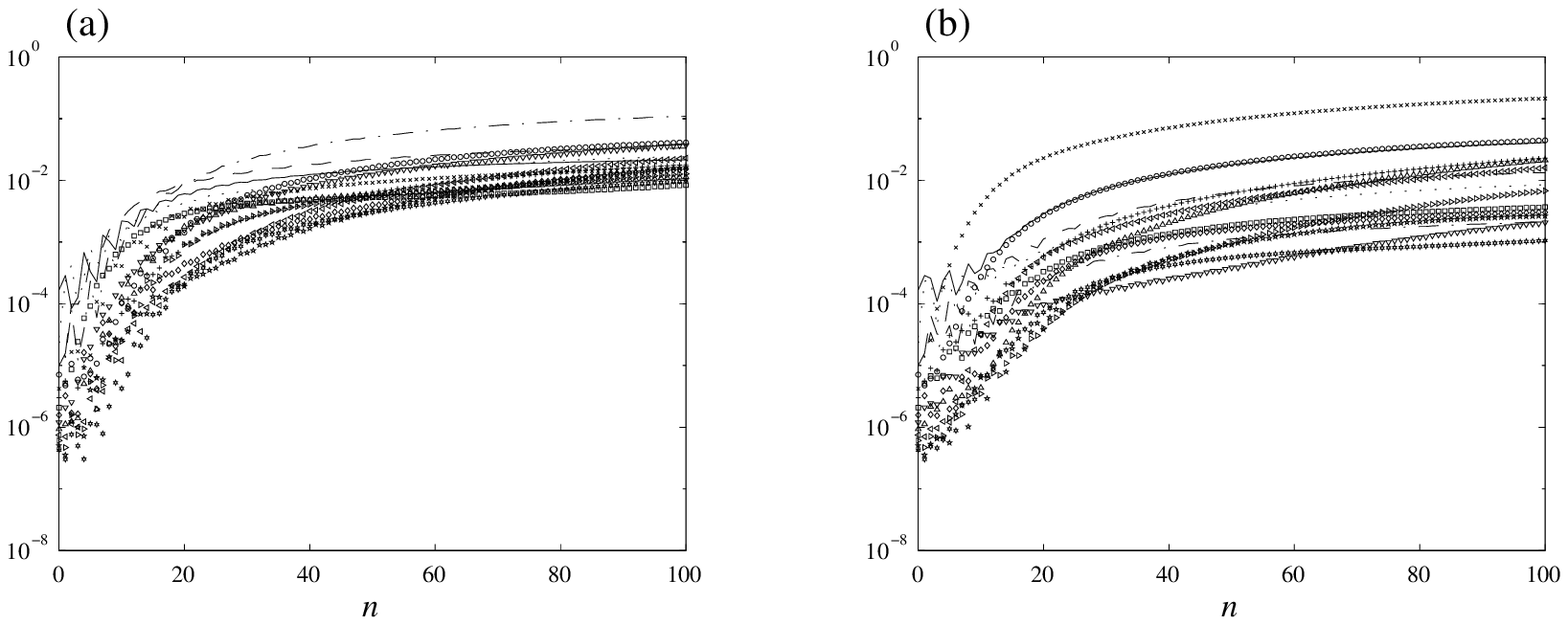,width=170mm}  
\end{center}
\caption{As for Fig.\ref{vkkoneone}, except that
$\eta'=2$, $\upsilon=1$. 
In (a) the leading term is  for $k=3$, in (b)  for $k=6$.}
\label{vkktwoone}
\end{figure}

\begin{figure}
\begin{center}
\epsfig{file=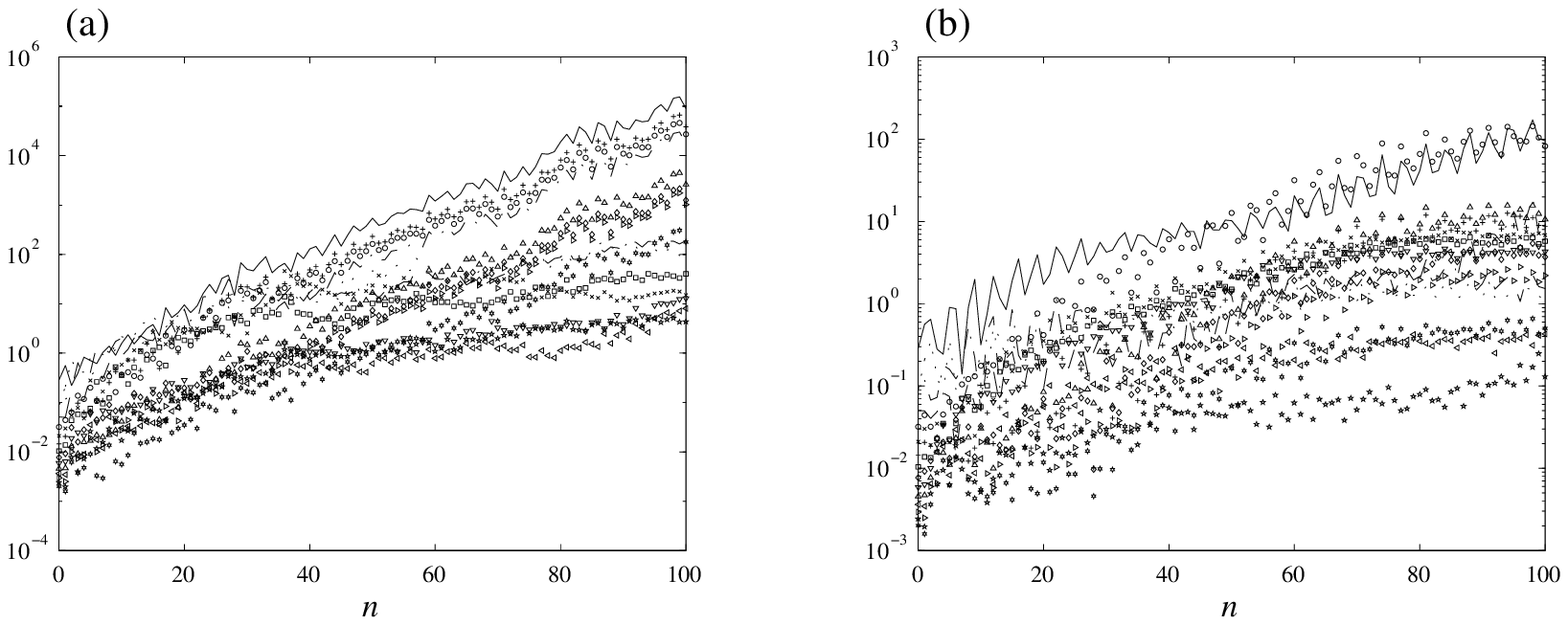,width=170mm}  
\end{center}
\caption{As for Fig.\ref{vkkoneone}, except that
$\eta'=1$, $\upsilon=10$. 
In (a) the leading term is  for $k=1$, in (b)  for $k=1,5$.}
\label{vkkoneten}
\end{figure}

\begin{figure}
\begin{center}
\epsfig{file=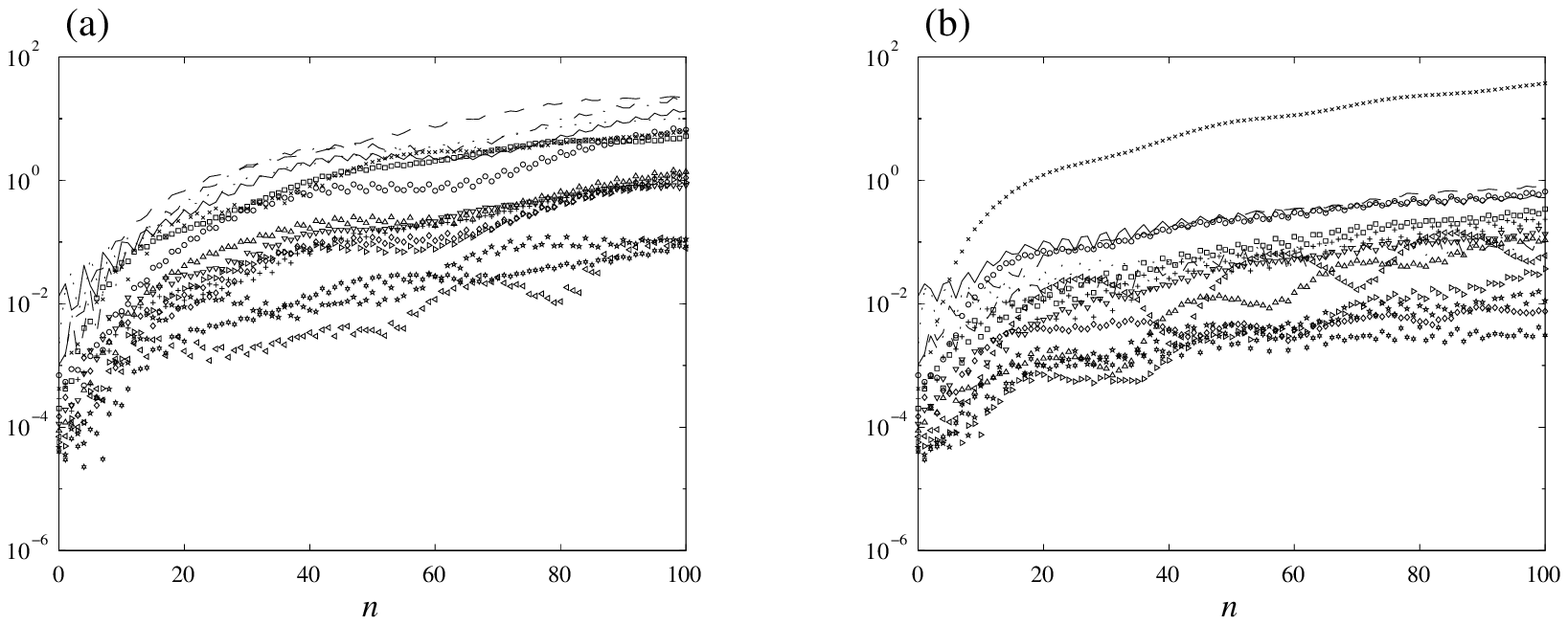,width=170mm}  
\end{center}
\caption{As for Fig.\ref{vkkoneone}, except that
$\eta'=2$, $\upsilon=10$. In (a) the leading term is for $k=4$, in (b) for
$k=6$.}
\label{vkktwoten}
\end{figure}

We integrated numerically Eq.~(\ref{Luveq}) for the first fifteen $u_{k}(x)$,
$v_{k}(x)$ pairs over the time span of 100 kicks, using a split operator method
described in some detail in Appendix~\ref{intLapp}, parallel to
numerical integration of the Gross-Pitaevskii equation, also using a split
operator method. Just before each kick each of the inner products 
$\langle v_{k}|v_{k}\rangle$ were determined, which are plotted against 
time in
Figs.~\ref{vkkoneone}--\ref{vkktwoten}, for
various parameter regimes we have already investigated the Gross-Pitaevskii
dynamics of. The ``stable'' and ``unstable'' initial conditions referred to 
are those of
the initial Gross-Pitaevskii wavefunction [which in turn determines the initial
conditions of each of the $u_{k}(x)$, $v_{k}(x)$ modes], and are exactly those
taken in the integrations of the Gross-Pitaevskii equation described in
Sec.~\ref{gpeint}. To reiterate, the data presented in the plots in this
section correspond exactly to the phase space plots presented in
Sec.~\ref{gpeint} for the appropriate values of $\upsilon$ and $\eta'$, with
regards to the initial condition. Figs.~\ref{vkkoneone},\ref{vkktwoone}
correspond to Figs.~\ref{uone},\ref{luone}, and
Figs.~\ref{vkkoneten},\ref{vkktwoten}
correspond to Figs.~\ref{uten},\ref{luten}.

In Fig.~(\ref{vkkoneone}), where $\eta'=1$ and $\upsilon=1$, we see a marked difference between the ``stable'' and
``unstable'' cases. In unstable case we see much greater growth of the 
$\langle v_{k}|v_{k}\rangle$. Interestingly, the $k=1$ mode in the stable
case does not on average seem to grow at all, instead undergoing quasiregular 
oscillations in time. The leading terms are also different; $k=1$ for the
unstable case, and $k=2$ in the unstable case.

Compared to Fig.~\ref{vkkoneone}, the ``stable'' and ``unstable'' cases shown in
Fig.~\ref{vkktwoone} (where the only difference is that $\eta'=2$), appear
comparatively similar. In particular there does not seem to be a great deal more
growth of the $\langle v_{k}|v_{k}\rangle$ in the unstable case when compared
to the stable case.

We see the same pattern repeated in Figs.~\ref{vkkoneten} and \ref{vkktwoten},
where $\upsilon$ is now 10. In Fig.~\ref{vkkoneten} the 
$\langle v_{k}|v_{k}\rangle$ very rapidly grow in the unstable case when
compared to the stable case, whereas in Fig.~\ref{vkktwoten}, where $\eta'=2$, the
difference is not nearly so marked (and in any case the growth of the 
$\langle v_{k}|v_{k}\rangle$ is generally less). 
This reflects in some sense the observed
Wigner function dynamics in Sec.~\ref{gpeint}, where there does not seem to be
such a strong qualitative difference between the ``unstable'' and ``stable''
cases where $\eta'=2$ for any value of $\upsilon$, in contrast to the cases 
where $\eta'=1$. 
One should bear in mind that although the dimensionless
nonlinearity strength $\upsilon$ is the same in both Figs.~\ref{vkkoneone} and
\ref{vkktwoone}, the actual repulsive interaction $Nu_{1d}$ is proportional 
to $\upsilon/\eta'^{3}$. 
One might argue then
that one would expect that there is 
generally less depletion from the wavefunction described by the 
Gross-Pitaevskii equation. The evolution of $\varphi(x,t)$ is also important
however:
$\upsilon/\eta'^{3}=1$ where $\upsilon=1$ and $\eta'=1$ is not that different from
$\upsilon/\eta'^{3}=1.25$ where $\upsilon=10$ and $\eta'=2$, but the evolutions of the 
$\langle v_{k}|v_{k}\rangle$ are.
There appears to be some correspondence between the Gross-Pitaevskii phase space
dynamics shown in Figs.~\ref{uone},\ref{uten} and the evolutions of the 
$\langle v_{k}|v_{k}\rangle$, in that when there is a significant difference
between the ``stable'' and ``unstable'' cases, this shows up in the dynamics of
the $\langle v_{k}|v_{k}\rangle$ corresponding to these different cases. Also a
more ``smooth'' phase space plot (as for $\eta'=2$ compared to $\eta'=1$ in
Figs.~\ref{uone},\ref{uten}) appears to correspond to a more ``smooth''
evolution of the $\langle v_{k}|v_{k}\rangle$
(Figs.~\ref{vkktwoone},\ref{vkktwoten} compared with 
Figs.~\ref{vkkoneone},\ref{vkkoneten}). As the equation describing the time
evolution of the $|u_{k}\rangle,|v_{k}\rangle$ pairs is essentially the same as
that describing the evolution of linearized orthogonal perturbations of the
Gross-Pitaevskii wavefunction \cite{castin}, this is not unexpected.

\subsection{Comparison with Experimental Parameters}
\begin{figure}
\begin{center}
\epsfig{file=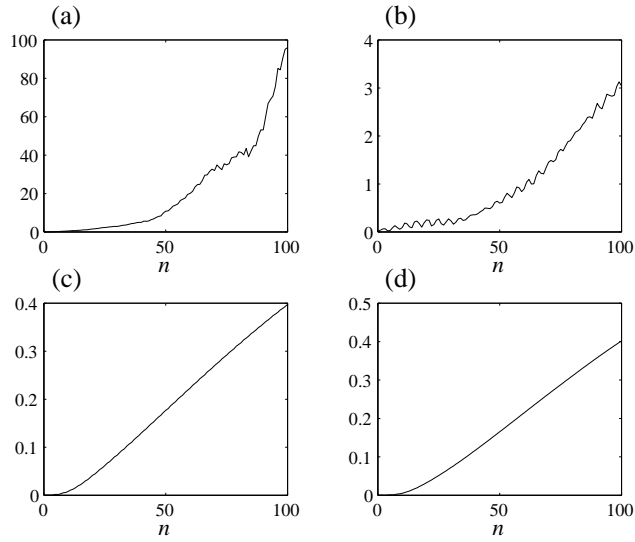,width=85mm}  
\end{center}
\caption{Plots of $\sum_{k=1}^{15}\langle v_{k}|v_{k}\rangle$ against the number
of kicks $n$, where
$\upsilon=1$, in the two cases of:
$\eta'=1$, for (a) unstable initial condition,
(b) stable initial condition;
$\eta'=2$, (c) unstable initial condition.
(d) stable initial condition.
}
\label{growthone}
\end{figure}

\begin{figure}
\begin{center}
\epsfig{file=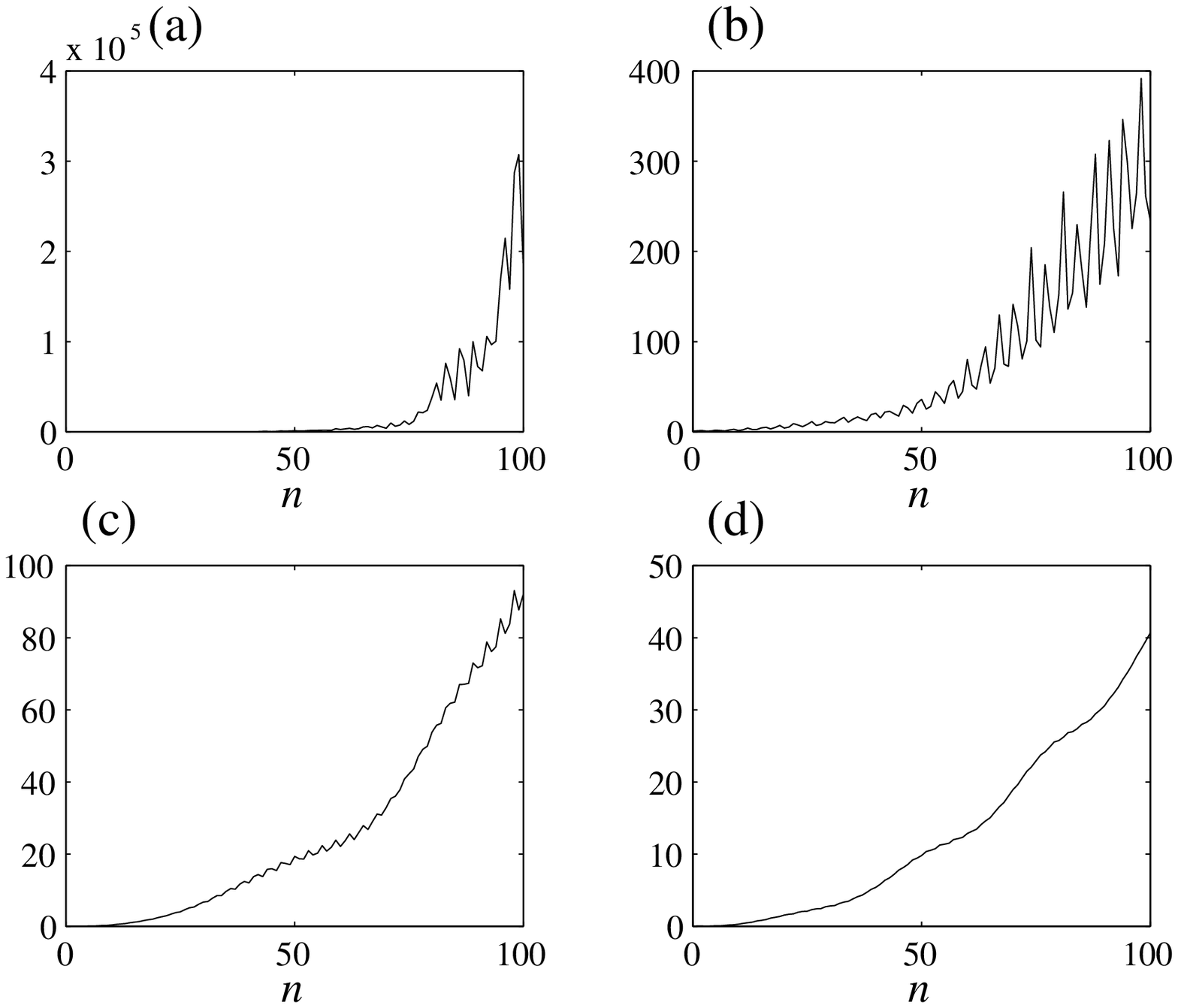,width=85mm}  
\end{center}
\caption{Corresponds exactly to Fig.~\ref{growthone}, except that
$\upsilon=10$.}
\label{growthten}
\end{figure}

We first examine our best estimate for $\langle \delta\hat{N}(t)\rangle$, which is
$\sum_{k=1}^{15}\langle v_{k}(t)|v_{k}(t)\rangle$, where $t$ is expressed as the
number of kicks. In Fig.~\ref{growthone} this
is plotted for each case where $\upsilon=1$ against the number of kicks, and in
Fig.~\ref{growthten} for $\upsilon=10$.
Interestingly, for $\upsilon=1$ and $\eta'=2$, total  growth appears to be almost 
exactly linear in time, after a short buildup period; as noted before, growth does
not appear to be that different when comparing the ``stable'' and ``unstable'' 
cases. For $\eta'=1$ however, there is a clear and substantial difference
between the two cases.

When $\upsilon$ is increased to 10, as shown in Fig.~\ref{growthten}, growth becomes
more erratic. We see that for the ``unstable'' case where $\eta'=1$, 
$\sum_{k=1}^{15}\langle v_{k}|v_{k}\rangle$ ends up being very large, making it
unlikely that an experiment for this parameter regime would follow Gross-Pitaevskii
dynamics. The general pattern observed in Fig.~\ref{growthone} is repeated here,
but with larger numbers. Note however, that the beginnings of a clear
differentiation
between the degree of growth for the ``stable'' and ``unstable'' cases when 
$\eta'=2$ appear to be occurring; in both cases growth is certainly not linear
with time.

Overall, our results can be interpreted as similar to those obtained in 
\cite{castindumdeplete} for the case of a time dependent harmonic potential. When one would 
expect classical chaotic behaviour, one observes rapid growth of the 
$\langle v_{k}|v_{k}\rangle$.

To examine the behaviour of a possible experimental realization of this scheme,
we consider Rubidium 87, which has an $s$ wave scattering length of
$a_{s}=5.1\times10^{-9}m$ \cite{rubidium}, 
and Sodium 23 ($a_{s}=2.75\times10^{-9}m$) \cite{sodium}.
Substituting Eq.~(\ref{etaprime}) into Eq.~(\ref{upsilonfull}), we can rewrite
$\upsilon$, so that
\begin{equation}
\upsilon=\sqrt{\frac{m}{\hbar\omega}}2N\omega_{r}a_{s}\eta'^{3}
\label{usefulupsilon}
\end{equation}
is expressed in terms of $\eta'$, which is more convenient for our purposes.
Using
Eq.~(\ref{usefulupsilon}), we get as a general relation for the number of
particles
$N=\lambda \sqrt{\omega}/\omega_{r}$, where
\begin{equation}
\lambda = 
\sqrt{\frac{\hbar}{m}}\frac{\upsilon}{2a_{s}\eta'^{3}}
\end{equation}
The values of $\lambda$ in units of $s^{-1/2}$ 
for the parameter regimes we have investigated are
summarized in Table~\ref{atomdata}. 

We let $\omega_{r}= 10\omega$, 
remembering that we should have $\omega_{r}$ significantly bigger than $\omega$,
we take this to be a reasonable minimum, bearing in mind that
 the values of the harmonic potential ground state
chemical potential $\mu$ lie 
between 0.55 and 3.11 in units of $\hbar\omega$, 
as shown in Table~\ref{atomdata}. We then get
$N=\nu/\sqrt{\omega_{r}}$, where $\nu=\lambda\sqrt{1/10}$. Numerical
values for $\nu$ in units of $s^{-1/2}$, where $\omega_{r}= 10\omega$ 
are also displayed in Table.~\ref{atomdata}. 
In principle this leaves us one free parameter to tweak; the smaller the radial
frequency, the larger $N$ can be, and the less significant the effect of the growth
of the number of particles not described by the Gross-Pitaevskii equation. This would
mean that we could reasonably expect to describe the dynamics of the particles
largely with the Gross-Pitaevskii equation, with small corrections accounted for by
Eq.~(\ref{Luveq}).

In practice trapping frequencies for alkali atoms such as Rubidium and Sodium  
lie between about 1 and 100 Hertz. The growth of $\sum_{k=1}^{15}\langle
v_{k}|v_{k}\rangle $ in the ``unstable'' case where $\upsilon=10$, $\eta'=1$ is thus 
far too high for this simplest interpretation of the real dynamics. The cases 
where $\eta'=2$ look more promising, and here in fact the interesting effect of
nonlinearity induced localization within phase space of the Gross-Pitaevskii 
wavefunction is even more pronounced. Also note that even for a small
nonlinearity of $\upsilon=1$, there is still a pronounced difference in the
Gross-Pitaevskii equation phase space dynamics (see Fig.~\ref{uone}) compared to
the case where there is no nonlinearity (Fig.~\ref{uzero}), for both $\eta'=1$
and $\eta'=2$, and here the numbers also seem more promising for the
nonlinearity induced localizing effect to be observed, corresponding to our
numerical integrations of the Gross-Pitaevskii equation.

\section{Conclusions}
We have derived explicitly an appropriate semiclassical limit for a general
cubic nonlinear Schr\"{o}dinger equation, or Gross-Pitaevskii equation, and find
it to be a Liouville type equation, with a term involving the density in
position space. We have shown how and why this differs 
from the hydrodynamic limit of the Gross-Pitaevskii equation. 
In particular, this derivation shows how an eccentric wavefunction 
$\varphi(x)$ can produce large deviations 
from this semiclassical limit, through higher order corrections involving 
derivatives of the density $\rho(x)=|\varphi(x)|^{2}$, in addition to 
effects due to an unusual potential. We have investigated
numerically a simple test system, the one-dimensional delta-kicked harmonic
oscillator, studying the dynamics of the Gross-Pitaevskii equation and the
appropriate Liouville type equation. We have found for moderate 
nonlinearity strengths that there is a localization effect explicitly 
due to interferences caused by the nonlinearity. We have outlined a possible
experimental implementation of such a system in a Bose-Einstein condensate
experiment, and have investigated numerically to what degree the
Gross-Pitaevskii equation describes correctly the dynamics of the bulk of the
particles for certain test cases. From this we have determined a lowest order 
estimate for the growth
in the number of non-condensate particles. We have found that for this system 
this depends strongly on the
parameter regime of $\eta'$ and $\upsilon$ under study, and that this seems to
correspond to the kinds of phase space dynamics observed in the Gross-Pitaevskii
equation. We have compared the numbers obtained with realistic experimental
parameters for condensates formed from sodium or rubidium atoms.

\section*{Acknowledgements}
We thank
J.~R.~Anglin, for helping clear up a number of points on the work
in Sec.~\ref{hydrowig}, M.~G.~Raizen, Th.~Busch, and K.~M.~Gheri, 
for discussions, and D.~A.~Steck for bringing reference  
\cite{borgonovi} to our attention.
We also thank the Austrian Science Foundation, and the European Union TMR network
ERBFMRX-CT96-0002.

\begin{appendix}
\section{Derivation of Wigner function dynamics}
\label{wignerapp}
\subsection{Definitions}
Defining the Wigner function for a pure state as
\begin{equation}
W(x,p) = \frac{1}{2\pi\hbar}\int_{-\infty}^{\infty}d\tau
e^{-ip\tau/\hbar} \varphi^{*}(x-\tau/2)\varphi(x+\tau/2),
\label{wignertwo}
\end{equation}
we take the time derivative
\begin{equation}
\frac{\partial}{\partial t}W(x,p) = 
\frac{\partial}{\partial t}W(x,p)_{\mbox{\scriptsize SP}}+\frac{\partial}{\partial t}W(x,p)_{\mbox{\scriptsize NL}},
\end{equation}
where we have split up the differential equation into a part which is governed
by the single particle linear dynamics (SP), and a part which is governed by
the nonlinearity (NL). 

\subsection{Single-particle dynamics}
The single particle dynamics are described by:
\begin{eqnarray}
\frac{\partial}{\partial t}W(x,p)_{\mbox{\scriptsize SP}}&=&
\frac{i}{2\pi\hbar^{2}}
\int_{-\infty}^{\infty}d\tau e^{-i\tau p/\hbar}
\left[
\langle \varphi |\hat{H}|x-\tau/2\rangle\langle x +\tau/2|\varphi\rangle 
-
\langle\varphi|x-\tau/2\rangle\langle x+\tau/2|\hat{H}|\varphi\rangle
\right].
\end{eqnarray}
The expansion we desire is exactly that used by
Zurek and Paz in investigating the quantum-classical boundary \cite{zurek}, and
is based on work originally carried out by Moyal\cite{moyal} 
and Wigner\cite{wigner}:
\begin{eqnarray}
\frac{\partial }{ \partial t}W(x,p)_{\mbox{\scriptsize SP}} &=&
\sum_{s=0}^{\infty}\frac{(-1)^{s}}{(2s+1)!} 
\left(\frac{\hbar}{2}\right)^{2s}
\frac{\partial^{2s+1}}{\partial x^{2s+1}}H
\frac{\partial^{2s+1}}{\partial p^{2s+1}}W
-\frac{\partial}{\partial p}H
\frac{\partial }{\partial x}W.
\label{appsingleparticle}
\end{eqnarray}

\subsection{Nonlinear Dynamics}
For a simple cubic nonlinearity $u|\varphi|^{2}\varphi$, we can express
$\partial W(x,p)_{\mbox{\scriptsize NL}}/\partial t$ as
\begin{eqnarray}
\frac{\partial}{\partial t}W(x,p)_{\mbox{\scriptsize NL}}&=&
\frac{iu}{2\pi\hbar^{2}}
\int_{-\infty}^{\infty}d\tau \left\{e^{-i\tau p/\hbar}
\int_{-\infty}^{\infty}dp'\left[
W(x-\tau/2,p')-W(x+\tau/2,p')
\right]
\int_{-\infty}^{\infty}dp''e^{i\tau p''/\hbar}W(x,p'')\right\}.
\label{nonlin}
\end{eqnarray}
We expand $W(x-\tau/2,p')-W(x+\tau/2,p')$
as a McLaurin series:
\begin{eqnarray}
\frac{\partial }{\partial t}W(x,p)_{\mbox{\scriptsize NL}}&=&-\frac{iu}{\pi\hbar^{2}}
\sum_{s=0}^{\infty}\frac{(1/2)^{2s+1}}{(2s+1)!}
\int_{-\infty}^{\infty}dp'\frac{\partial^{2s+1}}{\partial x^{2s+1}}W(x,p')
\int_{-\infty}^{\infty}d\tau e^{-i\tau p/\hbar}
\int_{-\infty}^{\infty}dp''\tau^{2s+1}e^{i\tau p''/\hbar}W(x,p'').
\end{eqnarray}
Using  the chain rule and Fourier's integral theorem, we arrive at
\begin{eqnarray}
\frac{\partial }{\partial t}W(x,p)_{\mbox{\scriptsize NL}}&=&
-i\frac{u}{\pi\hbar^{2}}
\sum_{s=0}^{\infty}\frac{(-\hbar/2i)^{2s+1}}{(2s+1)!}
\frac{\partial^{2s+1}}{\partial x^{2s+1}}
\left[\int_{-\infty}^{\infty}dp'W(x,p')\right]
\frac{\partial^{2s+1}}{\partial p^{2s+1}}W(x,p).
\label{expandnonlin}
\end{eqnarray}
\subsection{Combined Result}
Combining Eqs.~(\ref{appsingleparticle}) and (\ref{expandnonlin}), we get the
Wigner function dynamics to all orders in $\hbar$ of the cubic nonlinear
Schr\"{o}dinger equation with arbitrary potential, in one dimension 
\begin{eqnarray}
\frac{\partial}{\partial t}W&=&
\sum_{s=0}^{\infty}\frac{(-1)^{s}}{(2s+1)!}
\left(\frac{\hbar}{2}\right)^{2s}
\frac{\partial^{2s+1}}{\partial x^{2s+1}}
\left[H+u\rho\right]
\frac{\partial^{2s+1}}{\partial p^{2s+1}}W
-\frac{\partial }{\partial p}H\frac{\partial }{\partial x}W,
\label{appwignerexp}
\end{eqnarray}
which has as its semiclassical limit ($\hbar\rightarrow 0$) a Liouville-like
equation:
\begin{equation}
\frac{\partial}{\partial t}W=
\frac{\partial}{\partial x}
\left[H+u\rho\right]
\frac{\partial}{\partial p}W
-\frac{\partial }{\partial p}H
\frac{\partial }{\partial x}W,
\label{pseudohydro}
\end{equation}
where $\rho$ is the Wigner function integrated over $p$, as defined in
Eq.~(\ref{wigdens}).
This derivation can be easily generalized for other nonlinearities and to two and
three dimensions.

\section{Re-derivation of the hydrodynamic equations}
\label{hydroapp}
\subsection{Definitions}
The density $\rho$ has already been defined in terms of the 
Wigner function by Eq.~(\ref{wigdens}). The quantity $P$
is defined in terms of the Wigner function as
\begin{equation}
\rho P =  \int_{-\infty}^{\infty}dp p W.
\label{momentum}
\end{equation}

\subsection{Regaining the First Hydrodynamic Equation}
The equation of motion for $\rho$ is given by
\begin{eqnarray}
\frac{\partial }{\partial t}\rho &=&
\sum_{s=0}^{\infty}\frac{(-1)^{s}}{(2s+1)!}
\left(\frac{\hbar}{2}\right)^{2s}
\frac{\partial^{2s+1}}{\partial x^{2s+1}}\left[H+u\rho\right]
\int_{-\infty}^{\infty}dp\frac{\partial^{2s+1}}{\partial p^{2s+1}}W
-\int_{-\infty}^{\infty}dp\frac{\partial}{\partial x} W
\frac{\partial}{\partial p}H.
\end{eqnarray}
Due to the fact that $W(x,p)$ and all of its derivatives are equal to zero  at
$x=\pm\infty$, something we make frequent use of, 
this simplifies to the continuity equation
\begin{eqnarray}
\frac{\partial }{\partial t}\rho 
&=&-\frac{1}{m}\frac{\partial }{\partial x}(\rho P),
\label{appcontinuity}
\end{eqnarray}
using the definition of Eq.~(\ref{momentum}).

\subsection{Equations for Higher Order Moments}
We now turn to the equation of motion for $P$. We have, 
from Eq.~(\ref{momentum})
\begin{eqnarray}
\frac{\partial }{\partial t}P&=&\frac{1}{\rho}
\int_{-\infty}^{\infty}dp p 
\left\{
\sum_{s=0}^{\infty}\frac{(-1)^{s}}{(2s+1)!}
\left(\frac{\hbar}{2}\right)^{2s}
\frac{\partial^{2s+1}}{\partial x^{2s+1}}[V(x,t)+u\rho]
\frac{\partial^{2s+1} }{\partial p^{2s+1}}W-
\frac{p}{m}
\frac{\partial}{\partial x}W
\right\}
+\frac{P}{\rho m}
\frac{\partial }{\partial x}(\rho P).
\label{prehydro}
\end{eqnarray}
The integral of the Wigner function over $p$,
$\int_{-\infty}^{\infty} dp p \partial^{2s+1} W/\partial p^{2s+1}$, is 
equal to $\rho$ when $s=0$, and is otherwise
equal to zero. We therefore have
\begin{eqnarray}
\frac{\partial}{\partial t}P&=&-
\frac{\partial }{\partial x}[V(x,t)+u\rho]
-\frac{1}{\rho m}
\frac{\partial }{\partial x}\left(\int_{-\infty}^{\infty}dp p^{2}W\right)
+\frac{P}{\rho m}\frac{\partial}{\partial x}(\rho P).
\label{lasthydro}
\end{eqnarray}
Clearly Eq.~(\ref{appcontinuity}) and Eq.~(\ref{prehydro})
do not form a closed system of equations, due to the presence of the second
order moment $P_{2}(x)$, where
\begin{equation}
P_{n}(x) = \frac{1}{\rho(x)}\int_{-\infty}^{\infty}dp p^{n} W(x,p).
\label{mommoment}
\end{equation}
It is relatively simple to derive a chain of equations
of motion for all $P_{n}(x)$:
\begin{equation}
\frac{\partial }{\partial t}P_{n}(x)=
\frac{1}{\rho}\int_{-\infty}^{\infty} dp p^{n}\frac{\partial}{\partial t}W
-\frac{P_{n}(x)}{\rho}\frac{\partial}{\partial t}\rho.
\end{equation}
Substituting in Eqs.~(\ref{appwignerexp},\ref{appcontinuity}),
we get as the general form:
\begin{eqnarray}
\frac{\partial}{\partial t}P_{n}(x)&=&\frac{P_{n}(x)}{\rho m}
\frac{\partial}{\partial x}[\rho P(x)]
-\frac{1}{\rho m}\frac{\partial}{\partial x}[\rho P_{n+1}(x)]
-nP_{n-1}(x)\frac{\partial }{\partial x}[V(x,t)+u\rho]
\nonumber \\ &&
-\sum_{s=1}^{n-1}\left\{\frac{(\hbar/2)^{2s}n!}{(2s+1)![n-(s+1)]!}
P_{n-(s+1)}(x)\frac{\partial^{2s+1}}{\partial x^{2s+1}}[V(x,t)+u\rho]\right\}.
\label{motionmoment}
\end{eqnarray}
The system of equations Eqs.~(\ref{appcontinuity},\ref{motionmoment}), where $n$
ranges from $1$ to $\infty$, thus describes the full dynamics of the
Gross-Pitaevskii equation, Eq.~(\ref{gpe}) \cite{lill}. 

\subsection{Regaining the Second Hydrodynamic Equation}
We consider a set of solutions of the moments where 
$P_{n}(x)=P(x)^{n}$. 
Taking Eq.~(\ref{motionmoment}) and setting $\hbar=0$, i.e.
ignoring all quantum corrections, we substitute this solution in, which after
differentiation results in:
\begin{eqnarray}
nP(x)^{n-1}\frac{\partial}{\partial t}P(x)&=&
-\frac{nP(x)^{n}}{m}\frac{\partial}{\partial x}P(x)
-nP(x)^{n-1}\frac{\partial }{\partial x}[V(x,t)+u\rho],
\end{eqnarray}
where we can immediately carry out cancellations, to finally arrive at
\begin{eqnarray}
\frac{\partial}{\partial t}P(x)&=&-\frac{\partial }{\partial x}\left[
\frac{P(x)^{2}}{2m}
V(x,t)+u\rho\right],
\end{eqnarray}
which is the second hydrodynamic equation, Eq.~(\ref{momentumfield}). Thus
hydrodynamic equations describing dynamics in the hydrodynamic limit
\cite{review,hydro} are valid 
whenever $\hbar\rightarrow 0$ and $P_{n}(x)=P(x)^{n}$.
This condition can be expressed in terms of Liouville distributions as
\begin{equation}
\frac{1}{\rho}\int_{\infty}^{\infty} dp p^{n}(x)W(x,p)
=
\left[
\frac{1}{\rho}
\int_{\infty}^{\infty}
dp p W(x,p)
\right]^{n},
\end{equation}
which is in general fulfilled for $W(x,p)=\rho(x)\delta[p-p_{0}(x)]$, where
$p_{0}(x)$ is some single valued function of $x$.

\section{More Scaling}
\label{scalingapp}
As dimensionless parameters we have
$\eta$, $\kappa$, and 
$\upsilon$, defined in Eqs.~(\ref{eta},\ref{kappa},\ref{upsilon}), respectively.
We have as dimensionless coordinate and canonically conjugate momentum the
variables of Eqs.~(\ref{dimpos},\ref{dimmom}), and use the dimensionless time
$t_{h}=\omega t$.
Using this, we can write the dimensionless single
particle Hamiltonian
function as
\begin{eqnarray}
\tilde{H}&=&\frac{\tilde{p}^{2}}{2}+\tilde{V}(\tilde{x},t_{h})
 \label{dlessham}\\
\tilde{V}(\tilde{x},t_{h})&=&
\frac{\tilde{x}^{2}}{2}
+\frac{\kappa}{\sqrt{2}}\cos(\sqrt{2}\tilde{x})\sum_{n=-\infty}^{\infty}
\delta(t_{h}-n\tau_{h}),
\end{eqnarray}
the Gross-Pitaevskii equation, Eq.~(\ref{gpe}), as
\begin{equation}
i\frac{\partial}{\partial t_{h}}\tilde{\varphi}=
-\frac{\eta^{2}}{2}\frac{\partial^{2}}{\partial \tilde{x}^{2}}\tilde{\varphi}
+\frac{1}{\eta^{2}}\tilde{V}(\tilde{x},\tilde{t})\tilde{\varphi} 
+\frac{\upsilon}{\eta^{2}}|\tilde{\varphi}|^{2}\tilde{\varphi},
\end{equation}
and the equation of motion for the Wigner function as
\begin{eqnarray}
\frac{\partial}{\partial t_{h}}\tilde{W}&=&
\sum_{s=0}^{\infty}\frac{(-1)^{s}}{(2s+1)!}
\left(\frac{\eta^{2}}{2}\right)^{2s}
\frac{\partial^{2s+1}}{\partial \tilde{x}^{2s+1}}
\left[\tilde{H}+\upsilon\tilde{\rho}\right]
\frac{\partial^{2s+1}}{\partial \tilde{p}^{2s+1}}\tilde{W}
-\frac{\partial }{\partial \tilde{p}}\tilde{H}
\frac{\partial }{\partial \tilde{x}}\tilde{W}.
\label{etawigexp}
\end{eqnarray}
The wavefunction, Wigner function, and density, have been rescaled so that they are 
properly normalized: 
\begin{eqnarray}
\tilde{\varphi}&=&\sqrt{\sqrt{2}/k}\varphi,\\
\tilde{W}&=& \frac{2m\omega^{2}}{k^{2}}W,\\
\tilde{\rho}&=&\int_{-\infty}^{\infty}d\tilde{p}\tilde{W}.
\end{eqnarray} 
In the expansion shown in Eq.~(\ref{etawigexp})
it can clearly be seen that if $\eta$ is varied, then this is
completely independent of all other rescaled quantities. We see thus that
$\eta^{2}$ is an appropriate expansion parameter, and that the other
dimensionless parameters $\kappa$ and $\upsilon$ are correctly scaled to be
independent of the expansion parameter. 
If one takes only the zero order term in the sum, $\eta$ drops out
completely.

\section{Crystal Symmetry and Non-Localization}
\label{antilocalapp}
\subsection{Classical Background}
Consider the classical delta kicked harmonic oscillator described in
Eq.~\ref{dlessham}.
The symmetry properties of this system have been extensively investigated by
Zaslavsky and coworkers \cite{classharm,stochastic,web,symmetry}; we
recapitulate some of this to provide context.

One can determine a kick to kick mapping terms of 
$\alpha=(\tilde{x}+i\tilde{p})/\sqrt{2}$:
\begin{equation}
\alpha_{n+1}=\left[
\alpha_{n}+i\frac{\kappa}{\sqrt{2}}\sin(\alpha_{n}+\alpha_{n}^{*})
\right]e^{-i\omega\tau}.
\end{equation}
 If 
$\omega\tau=2\pi r/q$, then we can write the mapping after $q$ kicks as
\begin{equation}
\alpha_{n+q}=\alpha_{n}+i\frac{\kappa}{\sqrt{2}}
\sum_{k=0}^{q-1}\sin(\alpha_{n+k}+\alpha_{n+k}^{*})e^{i2\pi kr/q}.
\label{qmap}
\end{equation}
Keeping terms in $\kappa$ up to  first order only, we observe
an {\em approximate\/} rotational $q$ symmetry in phase space 
\cite{stochastic,symmetry};
if we substitute $\alpha_{n}$ with $\beta_{n}=\alpha_{n}e^{i2\pi l/q},
l\in{\Bbb{Z}}$,
we end up with $\beta_{n+q}=\alpha_{n+q}e^{i2\pi l/q}$.
There can also be
a translational symmetry in phase space, i.e. 
$
\beta_{n}=\alpha_{n} + \gamma \Rightarrow \beta_{n+q}=\alpha_{n+q} +
\gamma, \gamma \in {\Bbb{C}}.
$
Note that it is only possible 
to combine a rotational $q$ symmetry with translational symmetry when 
$q\in q_{c}=\{1,2,3,4,6\}$ \cite{tessellate}.

Translational symmetry demands
\begin{eqnarray}
\sum_{k=0}^{q-1}\sin(\alpha_{n+j}+\alpha_{n+j}^{*})e^{i2\pi kr/q}
&=&\sum_{k=0}^{q-1}\sin(\beta_{n+j}+\beta_{n+j}^{*})e^{i2\pi kr/q},
\end{eqnarray}
which in turn implies
$\beta_{n+j}+\beta_{n+k}^{*}=\alpha_{n+k}+\alpha_{n+k}^{*}+
2\pi\l_{k};\forall\; k,l_{k}\in{\Bbb{Z}}$. Thus, Eq.~(\ref{qmap}) for
$\beta_{n+q}$ can be
simplified to
\begin{eqnarray}
\beta_{n+q}&=&\alpha_{n}+\gamma_{0}+
i\frac{\kappa}{\sqrt{2}}
\sum_{k=0}^{q-1}\sin(\alpha_{n+k}+\alpha_{n+k}^{*}+\gamma_{k} +\gamma_{k}^{*})
e^{i2\pi kr/q},
\end{eqnarray}
where $\gamma_{k}=\gamma e^{-i2\pi kr/q}$. The condition for translational 
symmetry
is thus reduced to $\gamma_{k}+ \gamma_{k}^{*}= 2\pi l_{k}$, which implies
\begin{equation}
l_{k}=l_{0}\cos(2\pi kr/q)-i\frac{\gamma-\gamma^{*}}{2\pi}\sin(2\pi k r/q).
\label{classinteg}
\end{equation}
If we now let $k_{\pm}=q/2\pm m$ or $(q\pm m)/2$, depending on whether or not
$q$ is even, we get
\begin{eqnarray}
\cos(2\pi k_{+}r/q)&=&\frac{l_{k_{+}}+l_{k_{-}}}{2l_{0}}\in{\Bbb{Q}},
\label{gammaone}\\
i\frac{\gamma-\gamma^{*}}{\pi}\sin(2\pi k_{+} r/q)&=&l_{k_{-}}-l_{k_{+}}
\in{\Bbb{Z}}.
\label{gammatwo}
\end{eqnarray}
This implies that $\cos(2\pi/q)\in {\Bbb{Q}}$, and it is known that this can
only be true if $q\in q_{c}=\{1,2,3,4,6\}$ \cite{algebra}. This directly 
implies that $\cos(2\pi k r/q)\in {\Bbb{Q}}\;\forall \;k,r \in {\Bbb{Z}}$. 
There is is thus an {\em exact\/}
translational or {\em crystal\/} symmetry in phase space, 
for $q\in q_{c}$ only. There are an infinite number of
values of $\gamma$ for which this applies, 
determinable from Eqs.~(\ref{gammaone},\ref{gammatwo}).

\subsection{Quantum Expression}
Broadly following the treatment of Borgonovi and Rebuzzini \cite{borgonovi}, we
consider the unitary displacement operator 
$
D(\alpha)=e^{\alpha\hat{a}^{\dagger}-\alpha^{*}\hat{a}}
=e^{i(\varpi\hat{x}-\xi\hat{p})}
$ \cite{dan}.
The operators $\hat{a}^{\dagger}$ and $\hat{a}$ are the quantum harmonic
oscillator creation and annihilation operators, and
the operators $\hat{x}$, $\hat{p}$, are scaled in harmonic
units.
The displacement operator acting on a wavefunction is a quantum 
analogue to translating a classical point particle in phase space.
We now consider the Floquet operator
$
\hat{F}=e^{-i(\hat{a}^{\dagger}\hat{a}+1/2)\omega\tau}
e^{-i\kappa\cos[\eta(\hat{a}+\hat{a}^{\dagger})]/\sqrt{2}\eta^{2}}
$
and determine the commutation properties of it with the displacement operator.

Using elementary properties of coherent states \cite{dan}, it can be seen that
\begin{eqnarray}
D(\alpha)\hat{F}^{q}
&=&\prod_{k=0}^{q-1}\left\{
e^{-i(\hat{a}^{\dagger}\hat{a}+1/2)2\pi r/q}
e^{-i\kappa
\cos[\eta(\hat{a}+\hat{a}-\alpha_{k}-\alpha_{k}^{*})]/\sqrt{2}\eta^{2}}
\right\}D(\alpha),
\end{eqnarray}
where $\alpha_{k}=\alpha e^{i 2\pi k r/q}$. The product of Floquet operators
$\hat{F}^{q}$ corresponds to the mapping of Eq.~(\ref{qmap}) which we used to
investigate classical symmetry properties.

Thus, $D(\alpha)$ commutes with $\hat{F}^{q}$ if
$\eta(\alpha_{k}+\alpha_{k}^{*})=\sqrt{2}\eta\xi_{k}=2\pi l_{k},\quad l_{k}\in 
{\Bbb{Z}} \;\forall\; 
k$. Using this we arrive at, similarly to the 
derivation of Eq.~(\ref{classinteg}), 
\begin{equation}
l_{k}=l_{0}\cos(2\pi kr/q)-i\frac{(\alpha-\alpha^{*})\eta}{\sqrt{2}\pi}\sin(2\pi kr/q).
\end{equation}
Analogously to the classical case, 
$D(\alpha)$ commutes with $\hat{F}^{q}$ if and only if $q\in q_{c}$. 
This implies that for $q\in q_{c}$, the eigenstates of $\hat{F}^{q}$ are
invariant under certain displacements, of which there are an 
infinite number, and are thus extended. Localization is not expected to take 
place, similarly to the case of quantum
resonances in a delta-kicked rotor \cite{frasca,resonances}.

\section{Integration of the ${\cal L}$ Equation.}
\label{intLapp}
From \cite{castin}, we know that
\begin{equation}
i\hbar \frac{d}{dt}
\left(
\begin{array}{c}
|u_{k}(t)\rangle \\
|v_{k}(t)\rangle
\end{array}
\right)
=
{\cal L}
\left(
\begin{array}{c}
|u_{k}(t)\rangle \\
|v_{k}(t)\rangle
\end{array}
\right),
\end{equation}
and that the corresponding time evolution operator
\begin{equation}
{\cal U}(t)=
\left(
\begin{array}{cc}
\hat{Q}(t) & 0 \\
0 & \hat{Q}^{*}(t)
\end{array}
\right)
{\cal U}_{\mbox{\scriptsize GP}}(t)
\left(
\begin{array}{cc}
\hat{Q}(0) & 0 \\
0 & \hat{Q}^{*}(0)
\end{array}
\right).
\end{equation}
The operator ${\cal U}_{\mbox{\scriptsize GP}}(t)$
is the time evolution operator corresponding to 
${\cal L}_{\mbox{\scriptsize GP}}(t)$, given by
\begin{equation}
{\cal L}_{\mbox{\scriptsize GP}}(t)=
\left(
\begin{array}{cc}
V(\hat{x},t)+ 2u |\varphi(\hat{x},t)|^{2} + \hat{p}^{2}/2m 
& u \varphi(\hat{x},t)^{2} \\
-u\varphi(\hat{x},t)^{*2}
& -V(\hat{x},t)- 2u |\varphi(\hat{x},t)|^{2}-\hat{p}^{2}/2m
\end{array}
\right).
\end{equation}

In our case, the potential is that of the delta-kicked harmonic oscillator.
Integrating between kicks, we consider $V(\hat{x})$
 time independent. Note however, that ${\cal L}_{\mbox{\scriptsize GP}}(t)$ is still in principle
 time dependent through $\varphi(x,t)$. Thus,
taking very small time steps $\Delta t$, the evolution is given 
approximately by
\begin{equation}
\left(
\begin{array}{c}
|U_{k}(t+\Delta t)\rangle \\
|V_{k}(t+\Delta t)\rangle
\end{array}
\right)
\approx
e^{-i{\cal L}_{\mbox{\scriptsize GP}}(t)\Delta t/\hbar}
\left(
\begin{array}{c}
|U_{k}(t)\rangle \\
|V_{k}(t)\rangle
\end{array}
\right).
\end{equation}
The time evolution operator 
$e^{-i{\cal L}_{\mbox{\scriptsize GP}}(t)\Delta t/\hbar}$ can be split 
into position dependent and momentum
dependent parts, and the time evolution was then 
determined using a split operator method, of which there are
many variations \cite{bandrauk}. We set $|U_{k}(0)\rangle=|u_{k}(0)\rangle$
and $|V_{k}(0)\rangle=|v_{k}(0)\rangle$, and determined
$|u_{k}(t)\rangle$ and
$|v_{k}(t)\rangle$  
from
$|U_{k}(t)\rangle$ and
$|V_{k}(t)\rangle$
by projection, just before each kick .

The effect of a kick is given by:
\begin{equation}
\left(
\begin{array}{c}
u_{k}(x,t^{+}) \\
v_{k}(x,t^{+})
\end{array}
\right)
=
\left(
\begin{array}{c}
e^{-i\kappa\cos(\sqrt{2}\eta x)/\sqrt{2}\eta^{2}}u_{k}(x,t^{-}) \\
e^{i\kappa\cos(\sqrt{2}\eta x)/\sqrt{2}\eta^{2}}v_{k}(x,t^{-})
\end{array}
\right).
\end{equation}
In
Sec.~\ref{vkevol}, the procedure outlined
above was used, in conjunction with numerical integration of the Gross-Pitaevskii
equation, also by a split operator method with matching time steps. 

\end{appendix}

\begin{table}[htbp]
\begin{center}
\begin{tabular}{c|c|c||c|c||c|c}
\multicolumn{3}{c||}{\mbox{}} 
& 
\multicolumn{2}{c||}{Na$^{23}$} &
\multicolumn{2}{c}{Rb$^{87}$}
\\ 
\hline
$\upsilon$ & $\eta'$ & $\mu$  & $\lambda$ & $\nu$ & $\lambda$ & $\nu$ 
\\ 
\hline\hline
 1 &  1 & $0.87\hbar\omega$ & $9.55\times 10^{3}s^{-1/2}$ & $3.02\times 10^{3}s^{-1/2}$ & 
 $2.65\times 10^{3}s^{-1/2}$ & $8.38\times 10^{2}s^{-1/2}$\\
\cline{2-7}
  & 2 & $0.55\hbar\omega$ &$1.19\times 10^{3}s^{-1/2}$ & $3.77\times 10^{2}s^{-1/2}$ & 
  $3.31\times 10^{2}s^{-1/2}$ & $1.05\times 10^{2}s^{-1/2}$\\
\hline
 10 &  1 & $3.11\hbar\omega$ & $9.55\times 10^{4}s^{-1/2}$ & $3.02\times 10^{4}s^{-1/2}$ & 
 $2.65\times 10^{4}s^{-1/2}$ &  $8.38\times 10^{3}s^{-1/2}$\\
\cline{2-7}
  & 2 & $0.95\hbar\omega$ & $1.19\times 10^{4}s^{-1/2}$ & $3.77\times 10^{3}s^{-1/2}$ & 
  $3.31\times 10^{3}s^{-1/2}$ & $1.05\times 10^{3}s^{-1/2}$\\
\end{tabular}
\end{center}
\caption{Values of $\lambda$ and $\nu$ for Sodium 23 and Rubidium 87, when in the
parameter  regimes of $\upsilon$ and $\eta'$ under study. Also displayed are the
values of the numerically determined 
ground state chemical potential $\mu$ for the appropriate values
of  $\upsilon$ and $\eta'$, in units of $\hbar\omega$.}
\label{atomdata}
\end{table}


\begin{references}

\bibitem{reichl}
L.~E.~Reichl 
{\em The Transition to Chaos In Conservative Classical Systems: Quantum
Manifestations\/} (Springer-Verlag, New York 1992).

\bibitem{gutzwiller}
M.~C.~Gutzwiller 
{\em Chaos in Classical and Quantum Mechanics} (Springer-Verlag, Berlin 1990).

\bibitem{haake}
F.~Haake, 
{\em Quantum Signatures of Chaos\/} (Springer-Verlag, Berlin 1991).

\bibitem{peres}
A.~Peres, in {\em Quantum Chaos: Proceedings of the Adriatico Research
Conference on Quantum Chaos}, edited by 
H.~A.~Cerdeira, R.~Ramaswamy, M.~C.~Gutzwiller, and G.~Casati 
(World Scientific, Singapore, 1991); 
A.~Peres, 
{\em Quantum Theory: Concepts and Methods\/} (Kluwer Academic Publishers, 
Dordrecht 1993); 
for alternative treatments see also
R.~Schack and C.~M.~Caves
Phys.\ Rev.\ E {\bf 53}, 3257 (1996);
R.~Schack and C.~M.~Caves
Phys.\ Rev.\ E {\bf 53}, 3387 (1996); 
and
G.~Garcia de Polavieja,
Phys.\ Rev.\ A {\bf 57}, 3184 (1998).

\bibitem{zurek}
W.~H.~Zurek and J.~P.~Paz,
Phys.\ Rev.\ Lett.\ {\bf 72}, 2508 (1994);
W.~H.~Zurek and J.~P.~Paz,
Nuov.\ Cim.\ B {\bf 110}, 611 (1995).


\bibitem{nonlinchaos}
N.~Finlayson, K.~J.~Blow, L.~J.~Bernstein, and K.~W.~Delong,
Phys.\ Rev.\ A {\bf 48}, 3863 (1993);
F.~Benvenuto, G.~Casati, A.~Pikovsky, and D.~L.~Shepelyansky,
Phys.\ Rev.\ A {\bf 44}, R3423 (1994);
B.~M.~Herbst and M.~J.~Ablowitz,
Phys.\ Rev.\ Lett.\ {\bf 18}, 2065 (1989).

\bibitem{bosecond}
G.~Baym and C.~Pethick, 
Phys.\ Rev.\ Lett.\ {\bf 76}, 6 (1996);

\bibitem{review}
F.~Dalfovo, S.~Giorgini, L.~P.~Pitaevskii, and S.~Stringari,
Rev.\ Mod.\ Phys.\ {\bf 71}, 463 (1999).

\bibitem{nonlinearopt}
Y.~R.~Shen 
{\em Principles of Nonlinear Optics\/} (Wiley \& Sons, New York 1984);
R.~W.~Boyd 
{\em Nonlinear Optics\/} (Academic Press, San Diego 1992).

\bibitem{hydro}
S.~Stringari,
Phys.\ Rev.\ A {\bf 58}, 2385 (1998);
M.~Fliesser, A.~Csord\'{a}s, P.~Sz\'{e}pfalusy, and R.~Graham,
Phys.\ Rev.\ A {\bf 56}, R2533 (1997);
S.~Stringari,
Phys.\ Rev.\ Lett.\ {\bf 77}, 2360 (1996).

\bibitem{velocity}
It is more conventional to describe the hydrodynamic equations in terms of a
velocity field $V=P/m$, as in \cite{review,hydro}. We use $P$ because in this
way it is easier to have a consistent notation for the higher order moments that
appear in Sec.~\ref{wigner}. 

\bibitem{goldstein}
H.~Goldstein, 
{\em Classical Mechanics\/} (Addison-Wesley, Reading 1980).

\bibitem{lill}
Similar hydrodynamic expansions have been carried out for the linear
Schr\"{o}dinger equation, see:
J.~V.~Lill, M.~I.~Haftel, and G.~H.~Herling
Phys.\ Rev.\ A {\bf 39}, 5832 (1989);
J.~V.~Lill, M.~I.~Haftel, and G.~H.~Herling
J.\ Chem.\ Phys.\ {\bf 90}, 4940 (1989);
M.~Ploszajczak and M.~J.~Rhoades Brown, 
Phys.\ Rev.\ Lett.\ {\bf 55}, 147 (1985);
M.~Ploszajczak and M.~J.~Rhoades Brown, 
Phys.\ Rev.\ D {\bf 33}, 3686 (1986).

\bibitem{classharm}
G.~M.~Zaslavski\u{\i}, M.~Yu.~Zakharov, R.~Z.~Sagdeev, D.~A.~Usikov,
and A.~A.~Chernikov,
JETP Lett.\ {\bf 44}, 451 (1986);
G.~M.~Zaslavski\u{\i}, M.~Yu.~Zakharov, R.~Z.~Sagdeev, D.~A.~Usikov,
and A.~A.~Chernikov,
Sov.\ Phys.\ JETP {\bf 64}, 294 (1986);
A.~A.~Chernikov, R.~Z.~Sagdeev, D.~A.~Usikov, M.~Yu.~Zakharov, 
and G.~M.~Zaslavsky, 
Nature {\bf 326}, 559 (1987).

\bibitem{stochastic}
A.~A.~Chernikov, R.~Z.~Sagdeev, and G.~M.~Zaslavsky,
Physica D {\bf 33}, 65 (1988).

\bibitem{web}
V.~V.~Afanasiev, A.~A.~Chernikov, R.~Z.~Sagdeev, and G.~M.~Zaslavsky, 
Phys.\ Lett.\ A {\bf 144}, 229 (1990).

\bibitem{symmetry}
A.~A.~Chernikov, R.~Z.~Sagdeev, D.~A.~Usikov, and G.~M.~Zaslavsky, 
Computers Math.\ Applic.\ {\bf 17}, 17 (1989).

\bibitem{quantharm}
G.~P.~Berman, V.~Yu.~Rubaev, and G.~M.~Zaslavsky, 
Nonlinearity {\bf 4}, 543
(1991)

\bibitem{borgonovi}
F.~Borgonovi and L.~Rebuzzini,
Phys.\ Rev.\ E {\bf 52}, 2302 (1995). The possibility of extended eigenstates
was also investigated in \cite{quantharm}.

\bibitem{frasca}
M.~Frasca, 
Phys.\ Lett.\ A {\bf 231}, 344 (1997). 

\bibitem{hogg}
T.~Hogg and B.~A.~Huberman,
Phys.\ Rev.\ A {\bf 28}, 28 (1983).

\bibitem{arnold} 
V.~I.~Arnol'd, 
Sov.\ Math.\ Doklady {\bf 5}, 581 (1964), 
reprinted in {\em Hamiltonian Dynamical Systems}, edited by 
R.~S.~MacKay and J.~D.~Meiss
(Adam Kilger, Bristol 1986).

\bibitem{kam}
A.~N.~Kolmogorov, 
Dokl.\ Akad.\ Nauk.\ SSSR {\bf 98}, 527 (1954);
V.~I.~Arnol'd, 
Russ.\ Math.\ Survey {\bf 18}, 9, 85 (1963);
J.~Moser, 
Nachr.\ Akad.\ Wiss.\ G\"{o}ttingen II, Math.\ Phys.\ Kl.\ {\bf 18}, 1 (1962).

\bibitem{kohn}
W.~Kohn,
Phys.\ Rev.\ {\bf 4}, 1242 (1961);
S.~A.~Morgan, R.~J.~Ballagh, and K.~Burnett,
Phys.\ Rev.\ A {\bf 55}, 4338 (1997).

\bibitem{raizen}
F. L. Moore, J. C. Robinson, C. Bharucha, P. E. Williams, and M. G. Raizen, 
Phys.\ Rev.\ Lett.\ {\bf 73}, 2974 (1994);
J. C. Robinson, C. Bharucha, F. L. Moore, R. Jahnke, G. A. Georgakis, Q. Niu,
M. G. Raizen, and B.~Sundaram,
Phys.\ Rev.\ Lett.\ {\bf 74}, 3963 (1995);
J. C. Robinson, C. F. Bharucha, K. W. Madison, F. L. Moore, Bala Sundaram, S. R. Wilkinson,
and M. G. Raizen,
Phys.\ Rev.\ Lett.\ {\bf 76}, 3304 (1996);
B. G. Klappauf, W. H. Oskay, D. A. Steck, and M. G. Raizen
Phys.\ Rev.\ Lett.\ {\bf 81}, 1203 (1998);
B. G. Klappauf, W. H. Oskay, D. A. Steck, and M. G. Raizen
Phys.\ Rev.\ Lett.\ {\bf 81}, 4044 (1998). In the context of quantum chaos
significant experimental work has been carried out 
on microwave driven hydrogen \cite{misc}
and mesoscopic solid state systems \cite{fromhold}.

\bibitem{ionchaos}
S.~A.~Gardiner, J.~I.~Cirac, and P.~Zoller, 
Phys.\ Rev.\ Lett.\ {\bf 79}, 4790 (1997).

\bibitem{castin}
We use the formalism of Y.~Castin and R.~Dum, 
Phys.\ Rev.\ A {\bf 57}, 3008 (1998),
an analogous formalism is presented in
C.~W.~Gardiner,
Phys.\ Rev.\ A {\bf 56}, 1414 (1997);
 see also \cite{castindumdeplete}.

\bibitem{castindumdeplete}
Y.~Castin and R.~Dum, 
Phys.\ Rev.\ Lett.\ {\bf 79}, 3553 (1997).

\bibitem{fouriergrid}
C.~C.~Marston and G.~G.~Balint-Kurti,
J. Chem.\ Phys.\ {\bf 91}, 3571 (1989).

\bibitem{rubidium}
J.~L.~Roberts, N.~R.~Claussen, J.~P.~Burke, Jr., C.~H.~Greene, E.~A.~Cornell,
and C.~E.~Wieman,
Phys.\ Rev.\ Lett.\ {\bf 81}, 5109 (1998).

\bibitem{sodium}
J.~Stenger, S.~Inouye, M.~R.~Andrews, H.-J.~Miesner, D.~M.~Stamper-Kurn, and
W.~Ketterle,
Phys.\ Rev.\ Lett.\ {\bf 82}, 2422 (1999).

\bibitem{moyal}
J.~E.~Moyal, Proc.\ Cambridge Phil.\ Soc.\ {\bf 45}, 99 (1949).

\bibitem{wigner}
E.~P.~Wigner, Phys.\ Rev.\ {\bf 40}, 749 (1932).

\bibitem{tessellate}
L.~Fejes T\'{o}th
{\em Regular Figures\/}
(Pergamon, Oxford 1964).

\bibitem{algebra}
C.~W.~Curtis and I.~Reiner 
{\em Theory of Finite Groups and Associative Algebras\/} 
(Wiley-Interscience, New York 1962).

\bibitem{dan}
D.~F.~Walls and G.~J.~Milburn
{\em Quantum Optics\/} (Springer-Verlag, Berlin 1994).

\bibitem{resonances}
F.~M.~Izrailev and D.~L.~Shepelyanskii,
Sov.\ Phys.\ Dokl.\ {\bf 24}, 996 (1980);
F.~M.~Izrailev and D.~L.~Shepelyanskii,
Theor.\ Math.\ Phys.\ {\bf 43}, 553 (1980);
D.~R.~Grempel, S.~Fishman, and R.~E.~Prange,
Phys.\ Rev.\ Lett.\ {\bf 49}, 833 (1982);
D.~R.~Grempel, R.~E.~Prange, and S.~Fishman,
Phys.\ Rev.\ A {\bf 29}, 1639 (1984).

\bibitem{bandrauk}
See, for example
A.~D.~Bandrauk and H.~Shen,
J.\ Phys.\ A {\bf 27}, 7147 (1994), and references therein.

\bibitem{misc}
E.~Doron, U.~Smilansky, and A.~Frenkel, Phys.\ Rev.\ Lett.\ {\bf 65}, 3072
(1990);
J.~E.~Bayfield,
G.~Casati, I.~Guarneri, and D.~W.~Sokol,
Phys.\ Rev.\ Lett.\ {\bf 63}, 364 (1989);
E.~J.~Galvez, B.~E.~Sauer, L.~Moorman, P.~M.~Koch, and M.~Richards,
Phys.\ Rev.\ Lett.\ {\bf 61}, 2011 (1988); for a review see 
P.~M.~Koch and K.~A.~H.~van~Leeuwen,
Phys.\ Rep.\ {\bf 255}, 289 (1995).

\bibitem{fromhold}
P.~B.~Wilkinson,
T.~M.~Fromhold, L.~Eaves, F.~W.~Sheard, N.~Miura, and T.~Takamasu, 
Nature {\bf 380}, 608 (1996);
T.~M.~Fromhold,
P.~B.~Wilkinson, F.~W.~Sheard, L.~Eaves, J.~Miao, and G.~Edwards,
Phys.\ Rev.\ Lett.\ {\bf 75}, 1142 (1995).

\end{references}
\end{document}